\documentclass[twocolumn,showpacs,preprintnumbers,superscriptaddress,amsmath,amssymb,nofootinbib]{revtex4}
\usepackage{graphicx}% Include figure files
\usepackage{dcolumn}% Align table columns on decimal point
\usepackage{bm}% bold math

\usepackage[normalem]{ulem}  % \sout{old text} for strikeout
\usepackage[dvips]{color} % For blue in-text comments and additions

\renewcommand\sout{\bgroup \color{red} \ULdepth=-.5ex \ULset}

\newcommand{\Psfig}[2]{\includegraphics[width=#1]{#2}}

\newcommand{\SUN}[1]{\text{SU} ( #1 )}

\def\mev{\text{ MeV}}
\def\gev{\text{ GeV}}
\def\fm{\text{ fm}}
\def\Rho{\text{P}}

\begin{document}

\preprint{}

\title{Size measurement of dynamically generated resonances with finite boxes}

\author{Takayasu Sekihara} 
\altaffiliation[Present address: ]{Institute of Particle and Nuclear
  Studies, High Energy Accelerator Research Organization (KEK), 1-1,
  Oho, Ibaraki 305-0801,
  Japan.}
\affiliation{Department of Physics, Tokyo Institute of Technology,
  Tokyo 152-8551, Japan}

\author{Tetsuo Hyodo}
\affiliation{Department of Physics, Tokyo Institute of Technology,
  Tokyo 152-8551, Japan}

\date{\today}% It is always \today, today,

\begin{abstract}
  The structure of dynamically generated states is studied from a
  viewpoint of the finite volume effect.  We establish the relation
  between the mean distance between constituents inside a stable bound
  state and the finite volume mass shift. In a single-channel
  scattering model, this relation is shown to be valid for a bound
  state dominated by the two-body molecule component.  We generalize
  this method to the case of a quasi-bound state with finite width in
  coupled-channel scattering. We define the real-valued mean distance
  between constituents inside the resonance in a given closed channel
  using the response to the finite volume effect on the
  channel. Applying this method to physical resonances we find that
  $\Lambda (1405)$ and $f_{0}(980)$ are dominated by the $\bar{K} N$
  and $K \bar{K}$ scattering states, respectively, and that the
  distance between $\bar{K} N$ ($K \bar{K}$) inside $\Lambda (1405)$
  [$f_{0}(980)$] is $1.7$--$1.9 \fm$ ($2.6$--$3.0 \fm$).  The root
  mean squared radii of $\Lambda (1405)$ and $f_{0}(980)$ are also
  estimated from the mean distance between constituents.
\end{abstract}

% \keywords{Suggested keywords}%Use showkeys class option if keyword
\maketitle

\section{Introduction}

There are several hadrons which are expected to have some exotic
structures (exotic hadrons), and clarifying structures of these exotic
hadrons is one of the important tasks for the study of the strong
interactions~\cite{Nakamura:2010zzi}.  A classic example of exotic
hadrons is the hyperon resonance $\Lambda (1405)$, which is the
lightest baryon with spin-parity $J^{P} = 1/2^{-}$ although containing
one strange quark.  This resonance has been considered as a
quasi-bound state of the $\bar{K} N$ system~\cite{Dalitz:1960du,
  Dalitz:1967fp} owing to the strongly attractive $\bar{K} N$
interaction in the $I=0$ channel.  Another example is found in the
lightest scalar meson nonet [$f_{0}(600)=\sigma$, $\kappa (800)$,
$f_{0}(980)$, and $a_{0}(980)$], which exhibits an inverted spectrum
from the na\"{i}ve expectation with the $q \bar{q}$ assignment.  There
are several attempts to explain this anomaly, {\it e.g.}, multiquark
configurations for the scalar nonet~\cite{Jaffe:1976ig, Jaffe:1976ih}
and $K \bar{K}$ molecules for $f_{0}(980)$ and
$a_{0}(980)$~\cite{Weinstein:1982gc, Weinstein:1983gd}.  Recently
$\Lambda (1405)$ and the lightest scalar mesons are successfully
described by coupled-channel chiral dynamics (chiral unitary approach)
in meson-baryon~\cite{Kaiser:1995eg, Oset:1997it, Oller:2000fj,
  Lutz:2001yb,Hyodo:2011ur} and
meson-meson~\cite{Dobado:1993ha,Oller:1997ti, Oller:1998hw}
scatterings, respectively.

One of the characteristic features of exotic hadrons is the spatial
size, because one expects larger size of hadronic molecules than
ordinary hadrons. However, in general, candidates of exotic hadrons
are not in ground states but resonances with finite decay
width. Because of the decay process, mean squared radius of a
resonance is obtained as a complex number whose interpretation is not
straightforward~\cite{Sekihara:2008qk, Sekihara:2010uz}. To overcome
this difficulty, we recall the finite volume effect on bound
states. It has been shown in Refs.~\cite{Luscher:1985dn, Beane:2003da,
  Koma:2004wz, Sasaki:2006jn, Davoudi:2011md} that the binding energy
increases when a bound state of two particles is confined in a finite
box with periodic boundary condition. The reason is that the wave
function of the bound state in the box penetrates to the adjacent box
and hence the expectation value of the potential energy grows
negatively. This means that the finite volume effect is closely
related with the spatial structure of the bound state.

Motivated by these observations, in this study we aim at establishing
the relation between the finite volume effect and the spatial size of
both stable bound states and unstable resonance states, or more
precisely the mean distance between constituents inside the bound and
resonance states.  Firstly, we consider a stable bound state in
single-channel scattering where the mean distance between constituents
is well defined. We develop a method to evaluate the mean distance
from the finite volume effect, and examine its validity using a
dynamical scattering model. This method is straightforwardly
generalized to a bound state in coupled-channel scattering. In this
case, the size of the bound state is defined for each channel, which
can be estimated by the finite volume effect on the channel of
interest, with the other channels being unchanged.  Next we extend
this method to a resonance state in coupled-channel scattering, and
estimate the mean distance between constituents of the resonance in
closed channels.  As applications to physical states, we examine the
coupled-channel models for $\Lambda (1405)$ and the scalar mesons
$\sigma$, $f_{0}(980)$, and $a_{0}(980)$ to elucidate their
structures.

This paper is organized as follows. In Sec.~\ref{sec:formulation} we
formulate the size measurement of (quasi-)bound states using the
finite volume effect, and introduce a dynamical scattering model.  In
Sec.~\ref{sec:results} we examine the validity of our strategy using
the finite volume effect in the case of single-channel bound state,
and apply the method to physical hadron resonances.
Section~\ref{sec:conclusion} is devoted to the conclusion of this
study.

\section{Formulation}
\label{sec:formulation}

\subsection{Size measurement with finite volume effect}
\label{sec:size}

\begin{figure}[!t]
  \centering
  \begin{tabular*}{6cm}{@{\extracolsep{\fill}}c}
    \Psfig{6cm}{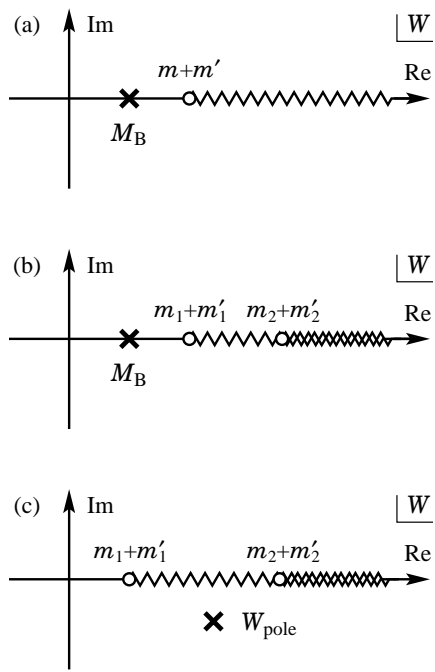} \\
  \end{tabular*}
  \caption{Schematic figure of (quasi-)bound state poles and continuum
    states in the complex plane of energy $W$.  (a) A bound state in
    single-channel scattering, (b) a bound state in coupled-channel
    scattering, and (c) a quasi-bound state in coupled-channel
    scattering.}
  \label{fig:1}
\end{figure}

Here we present the basic idea to determine the mean distance between
constituents inside the (quasi-)bound state from the mass shift due to
the finite volume effect.  The mean distance between constituents
inside the bound state is straightforwardly related to the size of the
system when the spatial size of constituents is negligible compared to
the distance between constituents.  Let us first consider the simplest
case: a bound state with mass $M_{\rm B}$ in the single-channel
scattering of particles with masses $m$ and $m^{\prime}\leq m$ [see
Fig.~\ref{fig:1} (a)]. In nonrelativistic quantum mechanics, the mean
squared distance of two particles in an weakly bound state can be read
off from the tail of the wave function as~\cite{Sekihara:2010uz}
\begin{align}
  \langle r^{2} \rangle_{\text{NR}}
  =& \frac{1}{4\mu B_{\rm{E}}},
  \label{eq:NRrms}
\end{align}
with the reduced mass $\mu = mm^{\prime}/(m+m^{\prime})$ and the
binding energy $B_{\rm{E}} = m+m^{\prime}-M_{\rm B}$. At first glance,
the mean squared distance for the bound state seems to be solely
determined by the mass of the bound state. However, it is implicitly
assumed in Eq.~\eqref{eq:NRrms} that the bound state is completely
described by the model space of the two-body scattering. In field
theory, in addition to the scattering state, there can be a ``bare
state'' contribution (elementarity) whose fraction in the physical
bound state is expressed by the wave function renormalization constant
$Z$~\cite{Weinberg:1965zz,PR136.B816}.\footnote{Strictly speaking, $Z$
  represents the contribution from those other than the present model
  space. Although this is not necessarily an elementary particle, for
  simplicity, we call $Z$ the ``bare state'' contribution in this
  paper.  Also we consider a point-like bare state which does not
  contribute to the size of the bound state.  This is in line with
  Ref.~\cite{Luscher:1985dn} where the finite volume effect is
  attributed to the modification of the loop of scattering states.}
The mean squared distance for the bound state, which stems from the
scattering state contribution, should then be given by subtracting the
bare state contribution as
\begin{align}
  \langle r^{2} \rangle
  =& \frac{1-Z}{4\mu B_{\rm{E}}} ,
  \label{eq:rms}
\end{align}
where the factor $1-Z$ is called the compositeness. As shown in
Refs.~\cite{Weinberg:1965zz,PR136.B816,Hyodo:2011qc}, the
compositeness is related with the coupling constant of the physical
bound state to the two-body scattering state $g$:
\begin{align}
  1-Z 
  =& - g^{2} 
  G^{\prime}(M_{\rm B}^{2}), 
  \label{eq:1mZ} \quad
  G^{\prime}(s)
  =\frac{d G(s)}{d s}  ,
\end{align}
where $G$ is the two-body loop integral as a function of squared
energy $s\equiv W^{2}$ to be specified below. Thus, the mean squared
distance for the bound state is expressed in terms of the mass $M_{\rm
  B}$ and its coupling to the scattering state $g$.

At this point we make use of the finite volume effect on the mass of a
bound state studied in Refs.~\cite{Luscher:1985dn, Beane:2003da,
  Koma:2004wz, Sasaki:2006jn, Davoudi:2011md}. When a bound state is
put in a periodic finite box of size $L$, the mass $M_{\rm B}$ is
shifted to $\tilde{M}_{\rm B}(L)$, and the mass shift $\Delta M_{\rm
  B}(L) \equiv \tilde{M}_{\rm B}(L) - M_{\rm B}$ is related to the
coupling constant $g$. As shown in Appendix~\ref{sec:appA}, the
leading contribution to the mass shift formula in the present case is
given by
\begin{align}
   \Delta M_{\rm B} ( L ) 
   =& -\frac{3 g^{2}}{8 \pi M_{\rm B}^{2} L}
   \exp \left [ - \bar{\mu} L \right ] +\mathcal{O}(e^{-\sqrt{2}\bar{\mu}L}), 
   \label{eq:mass-shift} \\
   \bar{\mu} =&\frac{\sqrt{- \lambda 
    (M_{\rm B}^{2}, \, m^{2}, \, m^{\prime 2})}}
    {2 M_{\rm B}} ,
    \label{eq:virtuality}
\end{align}
where $\lambda (x,\, y,\, z)=x^{2}+y^{2}+z^{2}-2xy-2yz-2zx$ is the
K\"{a}llen function. An important point is that the leading
contribution depends on $g^{2}$ and $M_{\rm B}$ in the infinite
volume. Namely, we can read off the physical coupling constant $g$
from the $L$ dependence of the mass of the bound
state. Equations~\eqref{eq:rms} and \eqref{eq:1mZ} show that the bound
state has large mean squared distance when the coupling $g$ is
large. This fact is intuitively understood in
Eq.~\eqref{eq:mass-shift} by the $g^{2}$ factor; a bound state with
large mean squared distance (and thus the large coupling $g$) has
strong finite volume effect. In other words, the structure of the
bound state is quantitatively reflected in the finite volume effect.
We also note that the factor $\bar{\mu}$ represents virtuality of the
constituent particles inside the bound state.

Equation~\eqref{eq:mass-shift} provides us an alternative strategy to
estimate the mean squared distance for a bound state. Suppose that we
are able to calculate the mass shift $\Delta M_{\rm B}(L)$ of the
bound state in finite volume at large $L$. In this case, using the
mass shift formula~\eqref{eq:mass-shift} up to the leading order, we
determine the coupling strength $g_{\rm FV}$. The mean squared
distance is then evaluated as
\begin{equation}
  \langle r^{2} \rangle _{\rm FV} 
  = \frac{-g_{\rm FV}^{2} G^{\prime}(M_{\rm B}^{2})}{4 \mu B_{\rm{E}}} .
  \label{eq:rmsfV}
\end{equation}
If the $L$ dependence of the mass shift is correctly fitted by
Eq.~\eqref{eq:mass-shift}, we expect $g_{\rm FV}\to g$ and $\langle
r^{2} \rangle _{\rm FV} \to \langle r^{2} \rangle$. In this way, the
mean squared distance for the bound state is related with the finite
volume mass shift at large $L$.

The virtue of this new approach will become clear when the argument is
extended to the quasi-bound state with finite width. To this end, we
begin with the case of a bound state in multichannel scattering as
shown in Fig.~\ref{fig:1} (b). Labeling the scattering states by the
suffix $i$, we can decompose the bound state wave function into the
bare state contribution ($Z$) and the contribution from the scattering
state in channel $i$ ($X_{i}$)~\cite{Hyodo:2011qc}, which are
normalized as
\begin{align}
  1
  =& Z + \sum_{i}X_{i}  ,
  \label{eq:1mZCoupled}\\
  X_{i}
  =&-g_{i}^{2}G_{i}^{\prime}(M_{\rm{B}}^{2}) .
  \label{eq:Xdef} 
\end{align}
In this case, the mean squared distance of the bound state in channel
$i$ is defined as
\begin{align}
  \langle r^{2} \rangle _{i}
  =& \frac{-g_{i}^{2}G_{i}^{\prime}(M_{\rm{B}}^{2})}{4\mu_{i} B_{\text{E},i}}  ,
  \label{eq:rmsCoupled}
\end{align}
with $g_{i}$ being the coupling constant to channel $i$, $B_{\text{E},
  i}=m_{i}+m^{\prime}_{i}-M_{\rm{B}}$ and
$\mu_{i}=m_{i}m^{\prime}_{i}/(m_{i}+m^{\prime}_{i})$. Conceptually,
$\langle r^{2} \rangle _{i}$ corresponds to the mean squared distance
measured by the probe current which exclusively couples to the
component in channel $i$.  Then the coupling constant $g_{i, \rm{FV}}$
can be extracted from the mass shift by putting only the channel $i$
in the finite box with size $L$ and keeping the other channels
unchanged. Substituting this coupling constant $g_{i, \rm{FV}}$ into
Eq.~\eqref{eq:rmsCoupled}, we obtain $\langle r^{2} \rangle
_{i,\rm{FV}}$.

We can further extend this argument to a quasi-bound state with finite
width. Consider a system with two coupled channels in which the higher
energy channel has a bound state when the transition potential is
switched off. The bound state acquires a decay width through the
channel coupling to the lower energy channel, which is called a
quasi-bound state or a Feshbach resonance [see
Fig.~\ref{fig:1} (c)]. In this case, the pole of the resonance
locates in the complex energy plane at total energy
$W=W_{\text{pole}}$. If the channel coupling is not strong, the
imaginary part of the pole position is small and we can identify the
real part as the ``mass'' of the state, $M_{\rm{B}}\equiv
\text{Re}[W_{\text{pole}}]$. Applying the same procedure, we determine
the coupling constant $g_{i,\rm{FV}}$ from the $L$ dependence of the
real part of the resonance pole when the channel $i$ is put in the
box. Substituting it in Eq~\eqref{eq:rmsCoupled}, we estimate the mean
squared distance for the quasi-bound state. The binding energy as well
as the loop integral are evaluated at this energy $M_{\rm{B}}\equiv
\text{Re}[W_{\text{pole}}]$.  Note that this is only applicable to the
closed channels, namely, the resonance pole should be located below
the threshold of channel $i$. If we put an open channel in the finite
box, the continuum state of that channel is discretized and we cannot
perform the analytic continuation to the complex energy plane.

It is important that this procedure gives a real-valued $\langle r^{2}
\rangle _{i}$ of the quasi-bound state, since the coupling extracted
from Eq.~\eqref{eq:mass-shift} is a real number. In general, it is
known that the mean squared radius~\cite{Sekihara:2008qk,
  Sekihara:2010uz} and compositeness~\cite{Hyodo:2011qc} become
complex in the case of resonances, which are difficult to
interpret. The strategy presented here can provide an alternative way
to investigate the structure of resonances.

Before closing this section, we comment on the modifications of the
formulation due to the finite size of the constituent particles.  As
discussed in Appendix~\ref{sec:appA}, the mass shift of the
constituent particles are in the higher order than the leading
contribution~\eqref{eq:mass-shift}, so this effect can be neglected at
least with sufficiently large $L$.  As a consequence, up to the
leading order, in which we are considering here, our formulation will
not be modified even if the constituent particles have their own
spatial structures.  Nevertheless, although the spatial structure of
the constituents does not affect the separation between constituents
in this study, mean squared radius of the whole system becomes larger
when the spatial structures of the constituents are taken into account
(see Appendix~\ref{sec:appB}).

In the following, we introduce a hadron scattering model
together with finite volume effect, in order to examine the size
measurement with finite box.

\subsection{Coupled-channel scattering model and finite volume effect}
\label{sec:infinite}

Here we formulate a model to describe stable bound states and unstable
resonance states along the line with Ref.~\cite{Oller:2000fj}.  We
prepare a coupled-channel interaction kernel $V_{ij}$ and evaluate the
scattering amplitude $T_{ij}$ by the Bethe-Salpeter equation in its
factorized form:
\begin{equation}
T_{ij} ( s ) = V_{ij} + \sum _{k} V_{ik} G_{k} T_{kj} 
= \sum _{k} {(1 - V G )^{-1}}_{ik} V_{kj} ,
\label{eq:BetheSalpeter}
\end{equation}
where indices $i$, $j$, and $k$ represent the scattering channels, $s
\equiv W^{2}$ is the squared center-of-mass energy of the scattering
system.  The explicit form of $V_{ij}$ will be given in the next
section. $G_{k}$ is the loop integral,
\begin{equation}
G_{k} ( s ) 
= i \int \frac{d ^{4} q}{(2 \pi)^{4}} 
\frac{1}{q^{2} - m_{k}^{2} + i \epsilon} 
\frac{1}{(P - q)^{2} - m_{k}^{\prime 2} + i \epsilon} , 
\end{equation}
with $m_{k}$, $m_{k}^{\prime}$, and $P^{\mu}=(W,\bm{0})$ being the
masses of the particles in channel $k$ and the four-momentum of the
two-body system, respectively.  Using the dimensional regularization,
one can rewrite the loop integral as
\begin{align}
G_{k} ( s ) 
= & \frac{1}{16 \pi ^{2}} \Bigg [ a_{k} (\mu _{\rm reg})
 + \ln \frac{m_{k}^{2}}{\mu _{\rm reg}^{2}} 
 + \frac{m_{k}^{\prime 2} - m_{k}^{2} + s}{2 s} 
\ln \frac{m_{k}^{\prime 2}}{m_{k}^{2}}
 \nonumber \\
& + \frac{\sqrt{\lambda _{k}}}{2 s}
\Big \{ \ln (s - m_{k}^{2} + m_{k}^{\prime 2} + \sqrt{\lambda _{k}}) 
 \nonumber \\
& \phantom{+ \frac{q_{k}}{\sqrt{s}}
\Big \{ }
+ \ln (s + m_{k}^{2} - m_{k}^{\prime 2} + \sqrt{\lambda _{k}}) 
\nonumber \\ 
& \phantom{+ \frac{q_{k}}{W}
\Big \{ }
- \ln (- s + m_{k}^{2} - m_{k}^{\prime 2} + \sqrt{\lambda _{k}})
\nonumber \\ 
& \phantom{+ \frac{q_{k}}{W}
\Big \{ }
- \ln (- s - m_{k}^{2} + m_{k}^{\prime 2} + \sqrt{\lambda _{k}}) \Big \}
\Bigg ] , 
\end{align}
with the regularization scale $\mu _{\rm reg}$, the subtraction
constant $a_{k}$, and $\lambda _{k} \equiv \lambda (s, \, m_{k}^{2},
\, m_{k}^{\prime 2})$.  We note that the regularization scale and the
subtraction constant are not independent, and the subtraction constant
is a single parameter of the loop function in each channel.

Bound states and resonance states appear as poles in the scattering
amplitude $T_{ij}$ as
\begin{equation}
T_{ij} ( s ) = \frac{g _{i} g_{j}}{s - s_{\rm pole}} 
+ T_{ij}^{\rm BG} ( s ) , 
\label{eq:amp_pole}
\end{equation}
where the background term $T_{ij}^{\rm BG}$ is chosen to make the
product $g_{i} g_{j}$ energy independent.  The constant $g_{i}$ can be
interpreted as the coupling strength of the state to the channel $i$.
The pole position $s_{\rm pole}$ is a solution of the equation,
\begin{equation}
\det ( 1 - V G ) = 0 , 
\label{eq:det}
\end{equation}
which is simplified as $V^{-1}=G$ in the single-channel case.  A
stable bound state is represented by a pole on the real axis of the
first Riemann sheet below the threshold, while an unstable resonance
state corresponds to a pole in the complex lower-half plane of the
second Riemann sheet above the threshold.

In this model, we have a relation, 
\begin{equation}
- \sum _{i, j} \left [ 
g_{i} \frac{d G_{i}}{d s} g_{i} \delta _{ij} 
+ g_{i} G_{i} \frac{d V_{ij}}{d s} G_{j} g_{j} 
\right ]_{s \to s_{\rm pole}} = 1 ,
\label{eq:Ward}
\end{equation}
which is the generalized Ward identity~\cite{Sekihara:2010uz,
  Hyodo:2011qc, Aceti:2012dd}. With Eq.~\eqref{eq:Xdef}, we can
identify the first term as the sum of the contributions from hadronic
composite states. It follows from Eq.~\eqref{eq:1mZCoupled} that the
bare pole contribution $Z$ is expressed by the second term as
\begin{equation}
Z = - \sum _{i, j} g_{i} G_{i} \frac{d V_{ij}}{d s} G_{j} g_{j} 
\Bigg | _{s \to s_{\rm pole}} . 
\label{eq:bare_pole}
\end{equation}
In Ref.~\cite{Hyodo:2011qc} $Z$ is shown to be exactly the bare pole
contribution for stable bound states, and thus for unstable resonance
states the system is expected to have less compositeness as $Z$
approaches unity.

Next we consider the finite volume effect in a spatial box with size
$L$.  In general, the finite volume effect appears as the discretized
momentum in the loop function~\cite{Luscher:1986pf}.  The finite
volume effect in the present model has been discussed in
Refs.~\cite{Doring:2011vk,MartinezTorres:2011pr} by using discretized
momentum loop function $\tilde{G}_{i}$,
\begin{align}
& \tilde{G}_{i} ( s ) 
= \frac{i}{L^{3}} \sum _{\bm{q}} \int \frac{dq^{0}}{2 \pi} 
\frac{1}{q^{2} - m_{i}^{2}} 
\frac{1}{(P - q)^{2} - m_{i}^{\prime 2}} , 
\nonumber 
\\
& \bm{q} = \frac{2 \pi \bm{n}}{L}, \quad 
\bm{n} \in \mathbb{Z}^{3} 
, 
\end{align}
instead of the loop integral $G_{i}$ in Eq.~\eqref{eq:BetheSalpeter}.
Here, we evaluate $\tilde{G}_{i}$ with the dimensional regularization
following Ref.~\cite{MartinezTorres:2011pr} by extracting
three-dimensional integral from $G_{i}$ and replace it with the
summation with discretized momentum, which results in,
\begin{align} &
  G_{i} (s) \to \tilde{G}_{i} (s) = G_{i} (s)
  \nonumber \\
  & + \lim _{q_{\rm max} \to \infty} \left ( \frac{1}{L^{3}} \sum
    _{\bm{q}}^{|\bm{q}|<q_{\rm max}} I_{i} (s,q) - \int _{q < q_{\rm
        max}} \frac{d^{3} q}{(2 \pi )^{3}} I_{i} (s,q) \right ) ,
\end{align}
with 
\begin{equation}
I_{i} ( s,q ) = \frac{1}{2 \omega _{i} \omega _{i}^{\prime}}
\frac{\omega _{i} + \omega _{i}^{\prime}}
{s - (\omega _{i} + \omega _{i}^{\prime})^{2}} , 
\end{equation}
\begin{equation}
\omega _{i} (q) = \sqrt{q^{2} + m_{i}^{2}} , 
\quad 
\omega _{i}^{\prime} (q) = \sqrt{q^{2} + m_{i}^{\prime 2}} . 
\end{equation}
It is known that, with finite cut-off $q_{\rm max}$, $\tilde{G}_{i}$
exhibits oscillations which gradually vanish as $q_{\rm max}$ goes to
infinity~\cite{MartinezTorres:2011pr}. This oscillation is caused by
the summation over the discretized momentum, which is not a continuous
function of $q_{\rm{max}}$. The absolute value of the integrand
$I_{i}(s,q)$ decreases for large $q$, so the discontinuity becomes
small with large $q_{\rm{max}}$. In order to make convergence with
respect to the oscillation, following
Ref.~\cite{MartinezTorres:2011pr}, we will take averaged value of
$\tilde{G}_{i}$ within range $q_{\rm max} \in [2 \gev , \, 4 \gev]$ in
the numerical calculation.

When we put all the channels in the finite box, momenta of the
scattering states above the threshold are also discretized, and the
eigenenergies are constrained by the pole
condition~\cite{Doring:2011vk}
\begin{equation}
\det ( 1 - V \tilde{G} ) = 0 ,
\label{eq:detFV}
\end{equation}
which again reduces to $V^{-1}=\tilde{G}$ in the single-channel
case. Note that for bound state poles below the threshold,
Eq.~\eqref{eq:det} and Eq.~\eqref{eq:detFV} are the same condition
with different loop function. For the application to the quasi-bound
state, we will use the loop function with channel $i$ in the finite
box as
\begin{equation}
G = \begin{pmatrix}
    G_{1} & & & &\\
    & G_{2} & & &\\
    & & \ddots & & \\
    & & & \tilde{G}_{i} & \\
    & & & & \ddots
    \end{pmatrix} .
\end{equation}
In this case, if the energy is smaller than the threshold of channel
$i$, $W< m_{i}+m^{\prime}_{i}$, the scattering amplitude
$T=(V^{-1}-G)^{-1}$ is a continuous function of $W$, and the resonance
pole can be searched for through the analytic continuation of the
amplitude in a usual manner.

\section{Results}
\label{sec:results}

\subsection{Size of bound states in single-channel scattering}
\label{sec:bound}

Now let us consider stable bound states in single-channel scattering
and see how they behave in the finite volume.  In Sec.~\ref{sec:size},
we have presented two methods to calculate the mean squared distance
between constituents inside the bound state.  The mean squared
distance in Eq.~\eqref{eq:rms} is obtained from the residue of the
pole, and that in Eq.~\eqref{eq:rmsfV} is evaluated by the finite
volume effect. In addition, the corresponding mean squared distance
can also be calculated by using the response to an external probe
current as shown in Ref.~\cite{Sekihara:2010uz}. Comparing the results
from different methods, we examine the validity of the size
estimation.

In this subsection, with the $\bar{K}N$ system in mind, we choose the
masses in the scattering state as $m=938.9 \mev$ and
$m^{\prime}=495.7 \mev$, respectively.  We use the natural
subtraction constant $a=-1.95$ with the regularization scale $\mu
_{\rm reg} = 630 \mev$, which is obtained to exclude explicit pole
contributions from the loop integral~\cite{Hyodo:2008xr}.  For the
interaction kernel $V$ we consider two types.  One is the constant
interaction,
\begin{equation}
V_{\rm I} = v_{0} , 
\end{equation}
with the energy independent parameter $v_{0}$ (case I).  The other 
interaction consists of a bare pole term,
\begin{equation}
V_{\rm II} = \frac{g_{0}^{2}}{s - s_{0}} , 
\end{equation}
with two parameters $g_{0}$ and $s_{0}$ which are constrained by
$g_{0}=\sqrt{s_{0}}$ for simplicity (case II).  The parameters $v_{0}$
and $g_{0}=\sqrt{s_{0}}$ are fixed so as to produce a bound state with
binding energy $B_{\rm E}=10 \mev$ in both cases, and as a result we
have $v_{0}=124.9$ and $g_{0}=\sqrt{s_{0}}=1.430 \gev$.

\begin{table}
  \caption{Properties of bound states in cases I and II. The results in column
    ``Pole'' are calculated from the residue of the pole in the models,
    and those in ``Finite Volume'' are obtained by using the finite volume
    effect.}
  \label{tab:1}
  \begin{ruledtabular}
    \begin{tabular*}{8.6cm}{@{\extracolsep{\fill}}ccc}
      Case I & Pole & Finite Volume
      \\
      \hline
      $g$ & $5.42 \gev$ 
      & $5.2$ -- $5.8 \gev$ 
      \\
      $Z$ & $0$ & $-0.14$ -- $0.09$ 
      \\
      $1-Z$ & $1$ & $\phantom{-} 0.91$ -- $1.14$ 
      \\
    \end{tabular*}
  \end{ruledtabular}
  \vspace{10pt}
  \begin{ruledtabular}
    \begin{tabular*}{8.6cm}{@{\extracolsep{\fill}}ccc}
      Case II & Pole & Finite Volume
      \\
      \hline
      $g$ & $1.38 \gev$ 
      & $1.7$ -- $2.6 \gev$ 
      \\
      $Z$ & $0.935$ 
      & $0.78$ -- $0.90$ 
      \\
      $1-Z$ & $0.065$ 
      & $0.10$ -- $0.22$
      \\
    \end{tabular*}
  \end{ruledtabular}
\end{table}

Properties of the bound states in two cases I and II are summarized in
the second column of Table~\ref{tab:1}.  The coupling constant $g$ is
calculated from the residue of the bound state pole as in
Eq.~\eqref{eq:amp_pole}, and the bare pole contribution $Z$ is
obtained by Eq.~\eqref{eq:bare_pole}.  As one can see from
Table~\ref{tab:1}, the bound state by the constant interaction in case
I has $Z=0$~\cite{Hyodo:2011qc}, which can be understood by
Eq.~\eqref{eq:bare_pole}. On the other hand, the bare pole potential
creates large elementarity $Z=0.935$ in case II.  Purely elementary
state with $Z=1$ can be obtained by taking the limit $g_{0}\to 0$ and
$s_{0} \to M_{\rm B}^{2}$ with fixed $B_{\rm E}$.

\begin{table}
  \caption{Mean distance between constituents inside bound states 
    in cases I and II with several methods. }
  \label{tab:2}
  \begin{ruledtabular}
    \begin{tabular*}{8.6cm}{@{\extracolsep{\fill}}cccc}
      & Eq.~\eqref{eq:rms} & Probe~\cite{Sekihara:2010uz} & Eq.~\eqref{eq:rmsfV}
      \\
      \hline
      Case I, $\sqrt{\langle r^{2} \rangle}$ [$\text{fm}$] 
      & $1.73 $ 
      & $1.86 $ 
      & $1.7$ -- $1.8$ 
      \\
      Case II,  $\sqrt{\langle r^{2} \rangle}$ [$\text{fm}$] 
      & $0.44 $ 
      & $0.48 $ 
      & $0.5$ -- $0.8$ 
      \\
    \end{tabular*}
  \end{ruledtabular}
\end{table}

With the obtained compositeness $1-Z$ and Eq.~\eqref{eq:rms}, we
calculate the mean distance for the bound state as shown in the second
column of Table~\ref{tab:2}.  Here we also calculate the mean distance
using probe method developed in Ref.~\cite{Sekihara:2010uz}, in which
the external probe current is coupled to the particles in the
scattering state and the mean squared distance is obtained from the
form factor. The results are shown in the third column of
Table~\ref{tab:2}.  Comparing two cases, we observe that the bound
state in case I has large separation between constituents
$\sqrt{\langle r^{2} \rangle}\sim 1.8 \fm$ compared to the hadronic
scale $\lesssim 0.8 \fm$, whereas in case II the separation for the
bound state is $\sqrt{\langle r^{2} \rangle}\sim 0.5 \fm$. This is
because only the two-particle cloud can contribute to the mean
distance.  These results indicate that the mean distance for the bound
state is not exclusively determined by its binding energy, and the
magnitude of the coupling constant is closely related with the
internal structure, as discussed in Sec.~\ref{sec:size}.

\begin{figure}[!t]
  \centering
  \begin{tabular*}{8.6cm}{@{\extracolsep{\fill}}c}
    \Psfig{8.6cm}{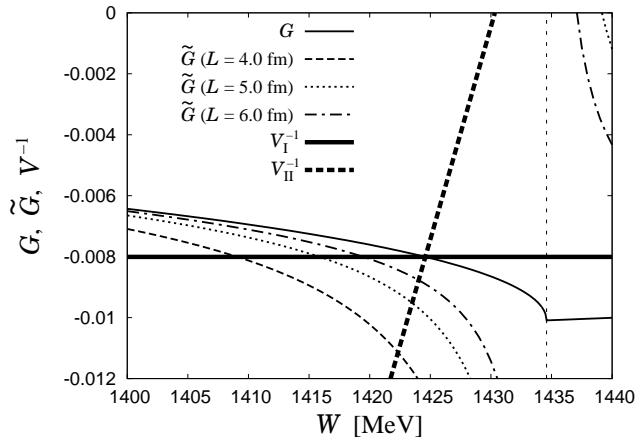} \\
  \end{tabular*}
  \caption{The loop integrals in infinite and finite volume, $G$ and
    $\tilde{G}$, and inverse of the interaction $V_{\rm I}$ and
    $V_{\rm II}$ as functions of $W \equiv \sqrt{s}$.  Vertical dotted
    line represents the threshold of the two-body system,
    $W=m+m^{\prime}$. }
  \label{fig:2}
\end{figure}

For later convenience, the loop integral $G$ and the inverse of the
interaction kernels $V_{\rm I}^{-1}$ and $V_{\rm II}^{-1}$ are plotted
as functions of $W \equiv \sqrt{s}$ in Fig.~\ref{fig:2}. Below the
threshold, the loop function $G$ is real. In this figure, the
intersection point of $G$ and $V^{-1}$ corresponds to the mass of the
bound state $M_{\rm B}$ according to Eq.~\eqref{eq:det}. In both
cases the intersection appears at $M_{\rm B}=1424.6 \mev$ with the
binding energy $B_{\rm E}=10 \mev$ with the adopted parameters.  An
important point to note here is that the energy dependence of two
interaction kernels is very different from each other.  While $V_{\rm
  I}^{-1}$ is completely flat, $V_{\rm II}^{-1}$ is almost vertical
with steep slope. In the limit of $g_{0}\to 0$ and $s_{0}\to
M_{\rm{B}}^{2}$, the slope becomes completely vertical.  This
difference of the interaction kernel will be crucial to the finite
volume effect on the bound states.

Then let us take into account the finite volume effect by replacing
the loop integral $G$ with that in finite volume $\tilde{G}$.
Behavior of $\tilde{G}$ is also plotted in Fig.~\ref{fig:2} with box
sizes $L=4.0$, $5.0$ and $6.0 \fm$.  Because of the pole
condition~\eqref{eq:detFV}, the mass of the bound state in finite
volume $\tilde{M}_{\rm{B}}(L)$ corresponds to the intersection point
of $\tilde{G}$ and $V^{-1}$.  From Fig.~\ref{fig:2}, one observes that
in both cases the mass of the bound state decreases when the box size
$L$ decreases.  However, $L$ dependence of the mass of the bound state
is quantitatively different in two cases.  The flat (steep) energy
dependence of $V_{\rm{I}}^{-1}$ ($V_{\rm{II}}^{-1}$) results in the
strong (mild) $L$ dependence of the bound state mass in finite volume.
Different $L$ dependence of the mass shift in two cases is understood
by this geometric argument.

\begin{figure}[!t]
  \centering
  \begin{tabular*}{8.6cm}{@{\extracolsep{\fill}}c}
    \Psfig{8.6cm}{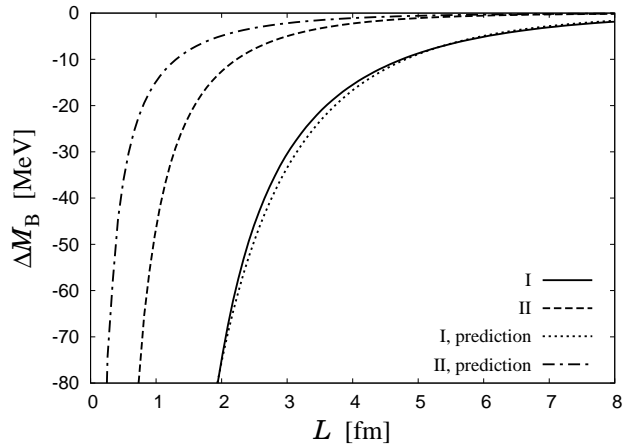} \\
  \end{tabular*}
  \caption{The mass shift of the bound states with two interactions
    $\Delta M_{\rm B}=\tilde{M}_{\rm B}-M_{\rm B}$ as a function of
    $L$.  The prediction by the mass shift
    formula~\eqref{eq:mass-shift} is also plotted. }
  \label{fig:3}
\end{figure}

To compare with the mass shift formula~\eqref{eq:mass-shift}, we plot
in Fig.~\ref{fig:3} the mass shift $\Delta M_{\rm
  B}(L)=\tilde{M}_{\rm{B}}(L)-M_{\rm{B}}$ as a function of $L$.  From
this figure, we observe the decrease of the mass for the smaller box
size $L$ in both cases I and II.  Furthermore, one can see the rapid
decrease of the mass in case I compared to that in case II.  This can
be interpreted as the consequence of the loose binding of the system
(large mean squared distance) in case I.

Using the coupling constant obtained from the pole residue, we can
predict the mass shift $\Delta M_{\rm B}$ by Eq.~\eqref{eq:mass-shift}
which is plotted in Fig.~\ref{fig:3}.  With large $L$, the
formula~\eqref{eq:mass-shift} well reproduces the mass shift, but some
deviation becomes evident in smaller $L$ region especially for the
case II. This means that higher order corrections on the mass shift
formula is necessary to describe finite volume effect of the bound
state. In fact, since the coupling $g$ is small in the case II, it is
reasonable that the higher order correction to the mass shift formula
is more important than the case I.

\begin{figure}[!t]
  \centering
  \begin{tabular*}{8.6cm}{@{\extracolsep{\fill}}c}
    \Psfig{8.6cm}{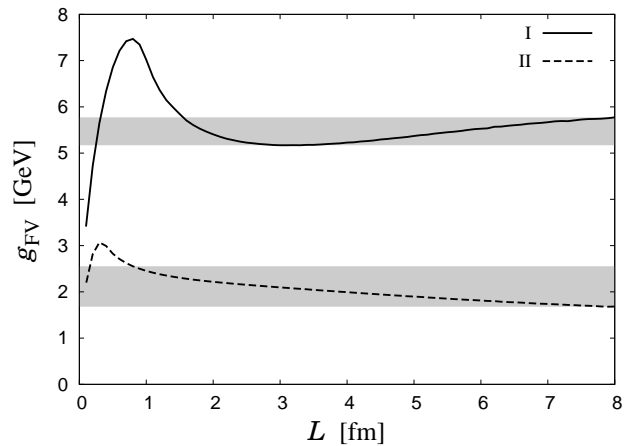} \\
  \end{tabular*}
  \caption{Coupling strength evaluated by Eq.~\eqref{eq:gFV} as a
    function of $L$.  The bands show the adopted values of $g_{\rm
      FV}$ in our study for both cases I and II. }
  \label{fig:4}
\end{figure}

Let us extract the bound state properties by using the finite volume
effect in the procedure of Sec.~\ref{sec:size}. Fitting the mass shift
by the formula~\eqref{eq:mass-shift}, we evaluate the coupling
strength $g_{\rm FV}$, the bare pole contribution $Z_{\rm FV}$, and
mean squared distance $\langle r^{2} \rangle _{\rm FV}$.  In this
study we take the following strategy to evaluate the coupling strength
$g_{\rm FV}$.  Namely, since the mass of the bound state is expected
to change according to Eq.~\eqref{eq:mass-shift} at the leading order,
the coupling strength $g_{\rm FV}$ can be extracted by the fraction of
the mass shift obtained in our model, $\Delta M_{\rm B}(L)$, and the
factor $-3 \exp ( - \bar{\mu} L )/ (8 \pi M_{\rm B}^{2} L)$ as,
\begin{equation}
  g_{\rm FV} = 
  \sqrt{\frac{\Delta M_{\rm B}(L)}
    {\displaystyle - \frac{3}{8 \pi M_{\rm B}^{2} L}
      \exp \left [ - \bar{\mu} L \right ]}} , 
\label{eq:gFV}
\end{equation}
with
\begin{equation}
  \bar{\mu} = \frac{\sqrt{- \lambda 
      (M_{\rm B}^{2}, \, m^{2}, \, m^{\prime 2})}}
  {2 M_{\rm B}} . 
\end{equation}
This $g_{\rm FV}$ depends on the box size $L$ especially in small $L$
region where the higher order contributions are not negligible.
Nevertheless, we expect that $g_{\rm FV}$ in Eq.~\eqref{eq:gFV}
becomes almost flat in the region where the mass shift is dominated by
the leading order contribution.  In Fig.~\ref{fig:4} we plot $g_{\rm
  FV}$ in Eq.~\eqref{eq:gFV} as a function of the box size $L$ for
both cases I and II.  From the figure, we can see that $g_{\rm FV}$ in
case I is fairly flat at $L \sim 3 \fm$, while it rapidly changes
below $\sim 2 \fm$ due to the higher order contributions to the mass
shift.  On the other hand, in case II $g_{\rm FV}$ increases without
flat regions as $L$ decreases down to $\sim 1 \fm$. Here in order to
determine the fairly flat region we make a criterion as follows.
Namely, according to Eqs.~\eqref{eq:mass-shift} and
\eqref{eq:amp_pole}, the typical scales of the box size $L$ and the
coupling strength $g$ are of the order of $1/\bar{\mu}$ and $M_{\rm
  B}=\sqrt{s_{\rm pole}}$, respectively.  Therefore, the typical
scales in Fig.~\ref{fig:4} are respectively $1/\bar{\mu}$ and $M_{\rm
  B}$ for the horizontal and vertical axes.  These characteristic
scales can make a model independent criterion $| d g_{\rm
  FV}/dL|<\bar{\mu} M_{\rm B}$ for the fit range to be fairly flat
$g_{\rm FV}(L)$.  This fit range corresponds to $L\in [1.9 \fm , \,
8.0 \fm]$ and $[0.8 \fm , \, 8.0 \fm]$ in case I and II, respectively,
with $\bar{\mu} M_{\rm B} = 0.58 \gev / \text{fm}$, and adopt $g_{\rm
  FV}$ in these ranges as the coupling strength from the finite volume
effect.  The adopted values of $g_{\rm FV}$ is shown as bands in
Fig.~\ref{fig:4} for both cases I and II.  The results are summarized
in Tables~\ref{tab:1} and~\ref{tab:2}.  As a result, we qualitatively
reproduce the structure of the bound state.  Especially the properties
of the bound state in case I are reproduced within $\sim 10 \%$
accuracy.  This indicates that the measurement of mean distance
between constituents with the finite volume effect is a powerful tool
to clarify the structure of bound states which have dynamical origin.

\subsection{Application to physical resonances}
\label{sec:resonance}

In the previous subsection we have developed a method to estimate the
separation between constituents inside the bound state by using the
finite volume effect.  One of the important features of our procedure
is the applicability to Feshbach resonance states with finite widths
as discussed in Sec.~\ref{sec:size}.  Furthermore, one can obtain
real-valued distance between constituents for the resonance states
with respect to a closed channel.  In this subsection we use this
method to discuss the structure of physical hadronic resonance states
from the finite volume.

Let us discuss $\Lambda (1405)$ in $\bar{K} N$-$\pi \Sigma$-$\eta
\Lambda$-$K \Xi$ coupled-channels and $\sigma$, $f_{0}(980)$, and
$a_{0}(980)$ scalar mesons in $\pi \pi$-$\pi \eta$-$K \bar{K}$
coupled-channels, assuming the isospin symmetry. These resonances have
been studied in chiral unitary approach~\cite{Kaiser:1995eg,
  Oset:1997it, Oller:2000fj, Lutz:2001yb, Hyodo:2011ur, Dobado:1993ha,
  Oller:1997ti, Oller:1998hw}, which is now elaborated using
next-to-leading order chiral interactions with recent experimental
data~\cite{Ikeda:2011pi,Ikeda:2012au,GomezNicola:2001as}.  To
concentrate on the size estimation with the finite volume effect, here
we utilize simplified models with leading order interactions as
follows.  For $\Lambda (1405)$ we employ the Weinberg-Tomozawa term as
the interaction kernel,
\begin{equation}
V_{ij} = - \frac{C_{ij}}{4 f^{2}}(2 W - M_{i} - M_{j}) 
\sqrt{4 M_{i} M_{j}}, 
\end{equation}
with $M_{i}$ being the baryon mass in channel $i$, $f$ the meson decay
constant, and $C_{ij}$ the Clebsch-Gordan coefficient which is
determined by the $\SUN{3}$ group structure of the interaction,
\begin{equation}
C_{ij} = 
\left ( 
\begin{array}{@{\,}cccc@{\,}}
  3 & - \sqrt{3/2} & 3 / \sqrt{2} & 0 \\
  - \sqrt{3/2} & 4 & 0 & \sqrt{3/2} \\
  3 / \sqrt{2} & 0 & 0 & - 3 / \sqrt{2} \\
  0 & \sqrt{3/2} & - 3 / \sqrt{2} & 3 
\end{array} 
\right ) , 
\end{equation}
where $i=1$, $2$, $3$, and $4$ denote the $\bar{K} N$, $\pi \Sigma$,
$\eta \Lambda$, and $K \Xi$ channels, respectively.  The meson decay
constant is $f=1.123 f_{\pi}$ with $f_{\pi}=93.0 \mev$.  The
subtraction constant is $a_{1}=-1.84$, $a_{2}=-2.00$, $a_{3}=-2.25$,
and $a_{4}=-2.67$ with the regularization scale $\mu _{\rm reg} = 630
\mev$ in all meson-baryon channels~\cite{Oset:2001cn}.  For the scalar
meson case, we take the lowest order $s$-wave meson-meson interaction
in chiral perturbation theory as the interaction kernel, namely,
\begin{equation}
V_{11} = \frac{m_{\pi}^{2} - 2 s}{2 f_{\pi}^{2}} , 
\quad 
V_{12} = V_{21} = - \frac{\sqrt{3} s}{4 f_{\pi}^{2}} , 
\quad 
V_{22} = - \frac{3 s}{4 f_{\pi}^{2}} , 
\end{equation}
for the $I=0$ channel with $i=1$ ($2$) for $\pi \pi$ ($K \bar{K}$)
channel, and
\begin{equation}
V_{11} = - \frac{m_{\pi}^{2}}{3 f_{\pi}^{2}} , 
\quad 
V_{12} = V_{21} = \frac{\sqrt{3/2}}{18 f_{\pi}^{2}} 
(9 s - m_{\pi}^{2} - 3 m_{\eta}^{2} - 8 m_{K}^{2}) , 
\end{equation}
\begin{equation}
V_{22} = - \frac{s}{4 f_{\pi}^{2}} , 
\end{equation}
for the $I=1$ channel with $i=1$ ($2$) for $\pi \eta$ ($K \bar{K}$)
channel.  Here we use the pion decay constant $f_{\pi}=93.0 \mev$.
The subtraction constant is fixed at $a=-1$ with the regularization
scale $\mu _{\rm reg}= 1.325 \gev$ in all meson-meson channels, which
corresponds to the three-dimensional cut-off $q_{\rm max}=1.092
\gev$~\cite{Oller:1998hw}.

\begin{table}
  \caption{Properties of resonances in $\bar{K}N$-$\pi\Sigma$-$\eta \Lambda$-$K
    \Xi$ scattering and $\pi\pi$-$\pi\eta$-$K\bar{K}$ scattering. Pole
    positions ($W_{\rm pole}$), coupling constants ($g$), and
    decomposition into scattering states ($X$) and bare pole
    contributions ($Z$) are shown.}
  \label{tab:3}
  \begin{ruledtabular}
    \begin{tabular*}{8.6cm}{@{\extracolsep{\fill}}ccc}
      & $\Lambda (1405)$, higher pole & $\Lambda (1405)$, lower pole
      \\
      \hline
      $W_{\rm pole}$ 
      & $1426.2-16.7 i \mev$ 
      & $1390.5-66.2 i \mev$ 
      \\
      $g_{\bar{K} N}$ 
      & $\phantom{-} 6.02 + 2.16i \gev$ 
      & $-2.69 + 4.15 i \gev$ 
      \\
      $g_{\pi \Sigma}$ 
      & $- 1.19 - 3.83 i \gev$ 
      & $\phantom{-} 6.25 - 4.10 i \gev$ 
      \\
      $g_{\eta \Lambda}$ 
      & $\phantom{-} 3.63 + 0.50 i \gev$ 
      & $\phantom{-} 0.03 + 1.98 i \gev$
      \\
      $g_{K \Xi}$ 
      & $- 0.35 - 0.93 i \gev$ 
      & $\phantom{-} 1.22 - 1.18 i \gev$ 
      \\
      % $\bar{K} N$
      $X_{\bar{K}N}$ 
      & $\phantom{-} 0.99 + 0.05 i $  
      & $- 0.21 - 0.13 i$
      \\
      % $\pi \Sigma$
      $X_{\pi\Sigma}$ 
      & $- 0.05 - 0.15 i $ 
      & $\phantom{-} 0.37 + 0.53 i $
      \\
      % $\eta \Lambda$
      $X_{\eta \Lambda}$ 
      & $\phantom{-} 0.05 + 0.01 i $ 
      & $- 0.01 + 0.00 i$
      \\
      % $K \Xi$
      $X_{K \Xi}$ 
      & $\phantom{-} 0.00 + 0.00 i $ 
      & $\phantom{-} 0.00 - 0.01 i $ 
      \\
      $Z$ 
      & $\phantom{-} 0.00 + 0.09 i $ 
      & $\phantom{-} 0.86 - 0.40 i $ 
      \\
    \end{tabular*}
  \end{ruledtabular}

  \vspace{10pt}
  \begin{ruledtabular}
    \begin{tabular*}{8.6cm}{@{\extracolsep{\fill}}cc}
      \multicolumn{2}{c}{$\sigma$}
      \\
      \hline 
      % Pole position
      $W_{\rm pole}$ & $471.3 - 181.0 i \mev$ 
      \\
      $g_{\pi \pi}$ & $\phantom{-} 1.84 - 2.31 i \gev$ 
      \\
      $g_{K \bar{K}}$ & $\phantom{-} 0.80 - 1.16 i \gev$ 
      \\
      % $\pi \pi$
      $X_{\pi\pi}$ & $-0.16 + 0.35 i $ 
      \\
      % $K \bar{K}$
      $X_{K\bar{K}}$ & $- 0.01 - 0.01 i $ 
      \\
      $Z$ & $\phantom{-} 1.17 - 0.34 i $ 
      \\
    \end{tabular*}
  \end{ruledtabular}
  \vspace{10pt}
  \begin{ruledtabular}
    \begin{tabular*}{8.6cm}{@{\extracolsep{\fill}}cc}
      \multicolumn{2}{c}{$f_{0}(980)$}
      \\
      \hline 
      % Pole position
      $W_{\rm pole}$ & $987.1 - 17.7 i \mev$ 
      \\
      $g_{\pi \pi}$ & $- 0.48 + 1.43 i \gev$ 
      \\
      $g_{K \bar{K}}$ & $\phantom{-} 3.91 + 1.32 i \gev$ 
      \\
      % $\pi \pi$
      $X_{\pi\pi}$ & $0.01 + 0.01 i $ 
      \\
      % $K \bar{K}$
      $X_{K\bar{K}}$ & $0.74 - 0.11 i $ 
      \\
      $Z$ & $0.25 + 0.10 i $ 
      \\
    \end{tabular*}
  \end{ruledtabular}
  \vspace{10pt}
  \begin{ruledtabular}
    \begin{tabular*}{8.6cm}{@{\extracolsep{\fill}}cc}
      \multicolumn{2}{c}{$a_{0}(980)$}
      \\
      \hline 
      % Pole position
      $W_{\rm pole}$ & $979.4 - 53.4 i \mev$ 
      \\
      $g_{\pi \eta}$ & $- 2.94 + 0.78 i \gev$ 
      \\
      $g_{K \bar{K}}$ & $\phantom{-} 4.58 + 0.48 i \gev$ 
      \\
      % $\pi \eta$
      $X_{\pi\eta}$ & $-0.06 + 0.10 i $ 
      \\
      % $K \bar{K}$
      $X_{K\bar{K}}$ & $\phantom{-} 0.38 - 0.29 i $ 
      \\
      $Z$ & $\phantom{-} 0.68 + 0.18 i $ 
      \\
    \end{tabular*}
  \end{ruledtabular}
\end{table}

With these interaction kernels, we obtain two resonance poles in the
meson-baryon scattering amplitude below the $\bar{K}N$ threshold, both
of which are associated with
$\Lambda(1405)$~\cite{Jido:2003cb,Hyodo:2007jq}. In meson-meson
scattering, we find two poles in $I=0$ and one in $I=1$ below the
$K\bar{K}$ threshold, which are interpreted as $\sigma$, $f_{0}(980)$,
and $a_{0}(980)$ mesons, respectively.  Properties of dynamically
generated resonances are summarized in Table~\ref{tab:3}.  The higher
pole of $\Lambda(1405)$ is expected to originate from the $\bar{K} N$
bound states~\cite{Hyodo:2007jq}, and in fact the magnitude of the
$\bar{K} N$ component $X_{\bar{K}N}$ is much larger than the others.
In the scalar meson case, we have $\sigma$ meson with very large width
$\sim 400 \mev$.  In the present setup, $f_{0}(980)$ is dominated by
the $K \bar{K}$ component whereas $a_{0}(980)$ shows large bare pole
contribution $Z$.  We note that the pole positions of $f_{0}(980)$ and
$a_{0}(980)$ depend on the cut-off for the loop integral and with
smaller $\mu _{\rm reg}$ they move above the $K \bar{K}$
threshold~\cite{Oller:1997ti}. In this case, the quasi-bound state
picture for $f_{0}(980)$ and $a_{0}(980)$ becomes unclear.  The input
models can be systematically improved within this approach using the
higher order chiral interaction and recent experimental
data~\cite{Ikeda:2011pi,Ikeda:2012au,GomezNicola:2001as}.

\begin{figure}[!t]
  \centering
  \begin{tabular*}{8.6cm}{@{\extracolsep{\fill}}c}
    \Psfig{8.6cm}{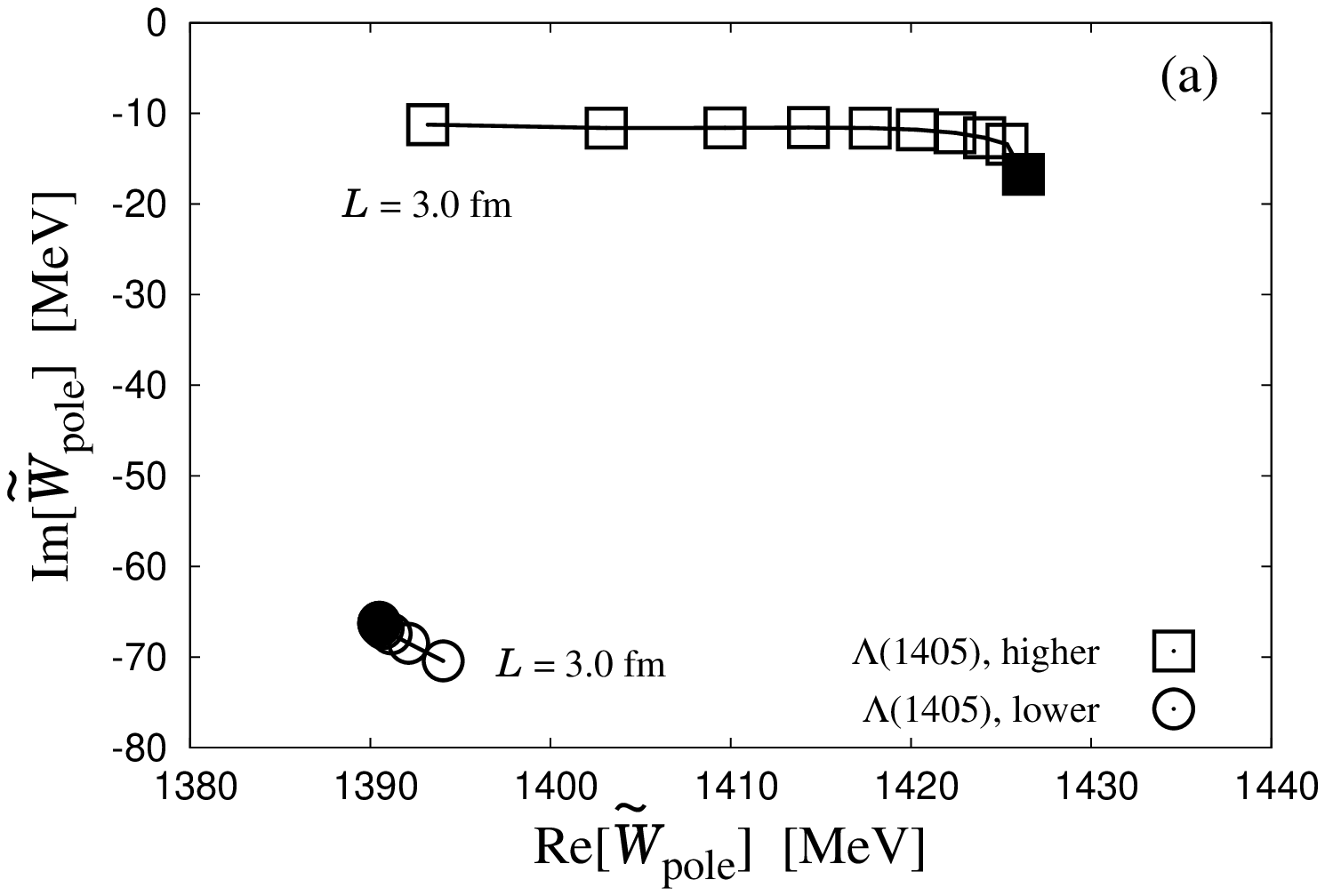} \\ 
    \Psfig{8.6cm}{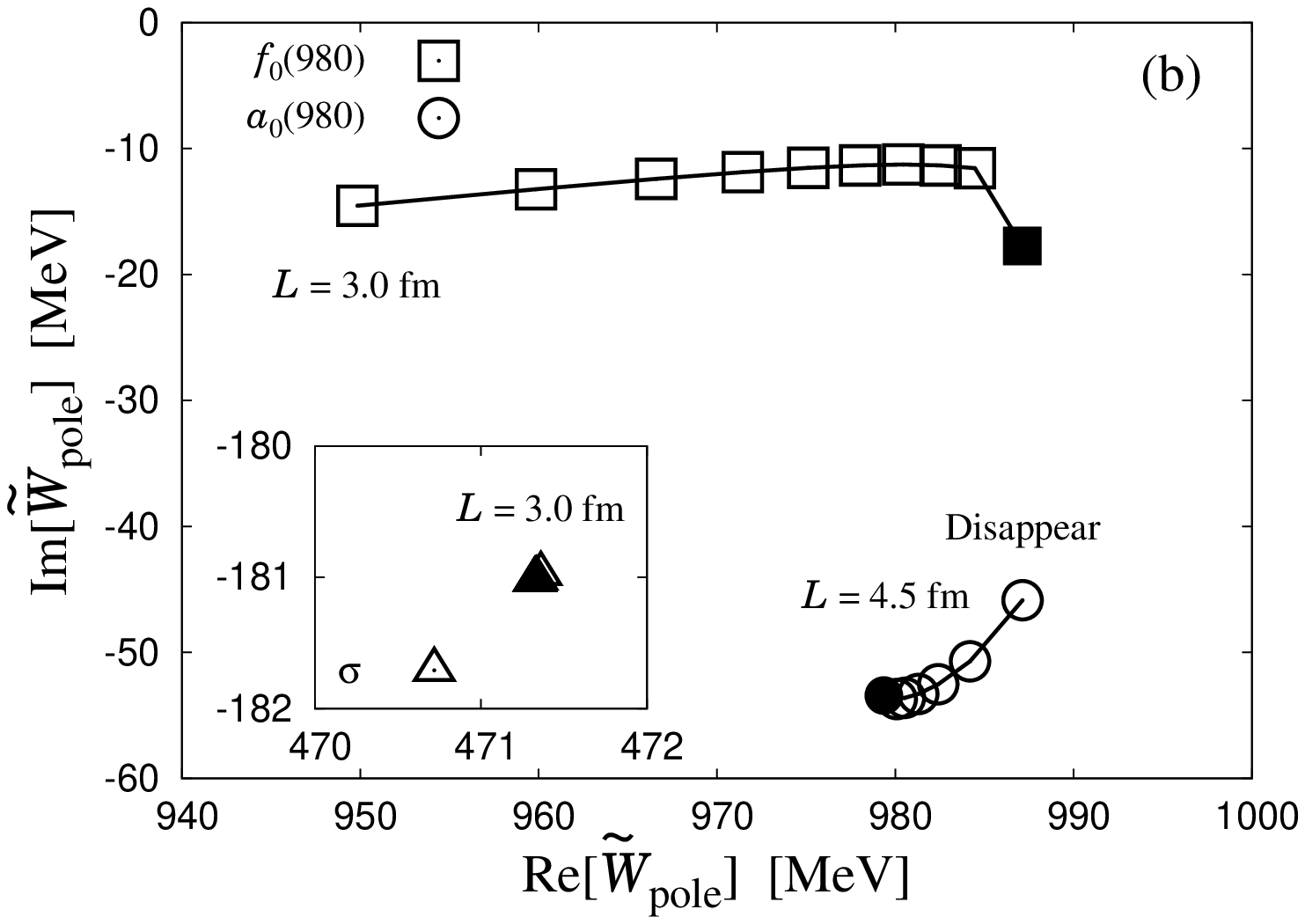} 
  \end{tabular*}
  \caption{Behavior of resonance pole positions for two $\Lambda
    (1405)$ (a) and three scalar mesons (b) with several box sizes
    $L$.  Here filled symbols indicate pole positions in infinite
    volume and open points are plotted in interval $0.5 \fm$ with
    respect to the box size $L$ from $L=3.0 \fm$ to $L=7.0 \fm$.  Note
    that $\sigma$ pole is quite stable with respect to the finite
    volume effect on the $K \bar{K}$ channel and the pole for
    $a_{0}(980)$ disappears for box sizes smaller than $4.5 \fm$.  }
  \label{fig:5}
\end{figure}

Then let us take into account the finite volume effect.  Since they
are the closed channels for all the poles considered here, we put
$\bar{K} N$ and $K \bar{K}$ channels into finite boxes with the
periodic boundary condition with other channels being unchanged.
Behavior of the resonance pole positions with respect to the box size
$L$ is shown in Fig.~\ref{fig:5}.  In the $\Lambda (1405)$ case
[Fig.~\ref{fig:5} (a)], the higher pole moves to lower energies when
the box size for the $\bar{K} N$ channel is reduced. On the other
hand, the lower pole stays around the original pole position even if
the finite volume effect on the $\bar{K} N$ channel is taken into
account.  This indicates that the higher pole is largely affected by
the modification of the $\bar{K}N$ loop and supports the scenario
that this pole originates from the $\bar{K} N$ bound state.  In the
scalar meson sector [Fig.~\ref{fig:5} (b)], $\sigma$ and $a_{0}(980)$
do not follow the expected mass shift formula; $\sigma$ is quite
stable with respect to the finite volume effect on the $K \bar{K}$
channel and the shift of the pole position is less than $1$ MeV. The
$a_{0}(980)$ pole disappears for box sizes smaller than $4.5 \fm$.  On
the other hand, the pole position of $f_{0}(980)$ shows strong $L$
dependence and moves to lower energies for smaller box size $L$.  This
implies large $K \bar{K}$ component inside $f_{0}(980)$, which is not
prominent for $\sigma$ and $a_{0}(980)$.

\begin{figure}[!t]
  \centering
  \begin{tabular*}{8.6cm}{@{\extracolsep{\fill}}c}
    \Psfig{8.6cm}{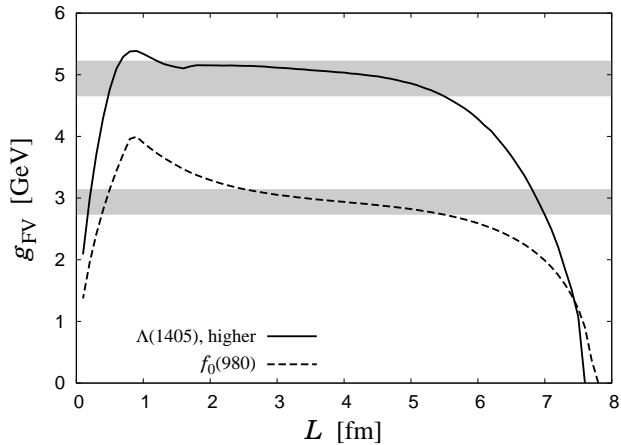} \\
  \end{tabular*}
  \caption{Coupling strength evaluated by Eq.~\eqref{eq:zgFV} as a
    function of $L$.  The bands show the adopted values of $g_{\rm
      FV}$ in our study for higher pole of $\Lambda (1405)$ and
    $f_{0}(980)$. }
  \label{fig:6}
\end{figure}

We next estimate the separation between constituents inside
dynamically generated resonances with the procedure developed in
Sec.~\ref{sec:size}.  In our approach, since we expect a downward
shift of the real part of the pole position in finite volume for
dynamically generated resonances, we firstly identify the real part of
the pole position as the mass of the state, and then estimate
uncertainties coming from the choice of the mass for the
resonances. However, our procedure is valid only when the resonance
originates from a bound state. In fact, the poles for $\sigma$,
$a_{0}$, and the lower energy pole of $\Lambda(1405)$ do not exhibit
the downward mass shift in finite volume. We then conclude that these
states are not dominated by the $\bar{K}N$ nor $K\bar{K}$ component,
in agreement with the results in Table~\ref{tab:3}. Therefore, we here
consider the properties of the higher pole of $\Lambda (1405)$ and
$f_{0}(980)$ resonance with respect to the $\bar{K} N$ and $K \bar{K}$
component, respectively.  We first fit the coupling strength
$g_{\bar{K}N,\text{FV}}$ ($g_{K\bar{K},\text{FV}}$) to the $L$
dependence of the real part of the pole position of the
$\Lambda(1405)$ [$f_{0}(980)$], and then evaluate the mean squared
distance between $\bar{K} N$ ($K \bar{K}$) in $\Lambda (1405)$
[$f_{0}(980)$].  As in the case of the bound state, we extract the
coupling strength $g_{\rm FV}$ by,
\begin{equation}
g_{\bar{K} N (K \bar{K}),\text{FV}} = 
\sqrt{\frac{\text{Re} [ \tilde{W}_{\rm pole}] - M_{\rm B}} 
{\displaystyle - \frac{3}{8 \pi M_{\rm B}^{2} L}
   \exp \left [ - \bar{\mu}_{\bar{K} N (K \bar{K})} L \right ]}} , 
\label{eq:zgFV}
\end{equation}
\begin{equation}
  \bar{\mu}_{\bar{K} N (K \bar{K})} =\frac{\sqrt{- \lambda 
      ( M_{\rm B}^{2}, \, m_{K}^{2}, \, m_{N (K)}^{2})}}
  {2 M_{\rm B}} ,
\end{equation}
for higher pole of $\Lambda (1405)$ [$f_{0}(980)$], and for the
resonance mass we take $M_{\rm B}=\text{Re}[W_{\rm
  pole}]$.\footnote{Note that the masses of $K$ and $\bar{K}$ are the
  same, but they are distinguishable by the strangeness quantum
  number.  We use the formula~\eqref{eq:mass-shift}, which differs by
  factor $2$ from the one for identical particles in
  Ref.~\cite{Luscher:1985dn}.}  Here $W_{\rm pole}$ and
$\tilde{W}_{\rm pole}$ are resonance pole position in the complex
energy plane in infinite and finite volume, respectively.  In
Fig.~\ref{fig:6} we plot $g_{\rm FV}$ in Eq.~\eqref{eq:zgFV} as a
function of the box size $L$.  In the figure, we observe rapid change
of $g_{\rm FV}$ also in large $L$ region.  This is because the pole of
the resonance states does not simply move downward in large $L$
region.  When $\text{Re} [\tilde{W}_{\rm pole}] = M_{\rm B}$ the
coupling strength $g_{\rm FV}$ defined in Eq.~\eqref{eq:zgFV}
vanishes, which takes place at $L\sim 7$--8 fm region in
Fig.~\ref{fig:6}.  For $\text{Re} [\tilde{W}_{\rm pole}] > M_{\rm B}$,
$g_{\rm FV}$ becomes pure imaginary.  Such upward shift of $\Lambda
(1405)$ and $f_{0}(980)$ is caused by the repulsion from the lower
energy pole [the lower pole of $\Lambda (1405)$ and $\sigma$] in the
complex energy plane.  For sufficiently small $L$, the downward
movement overcomes the repulsion and the mass shift follows
Eq.~\eqref{eq:mass-shift}.  In this case, we observe fairly flat
$g_{\rm FV}$ in range $\sim 4 \fm$ both for the higher pole of
$\Lambda (1405)$ and $f_{0}(980)$.  Hence, as in the bound state case,
we adopt $g_{\rm FV}$ in the region where $| d g_{\rm FV}/dL| <
\bar{\mu} _{\bar{K}N, K\bar{K}} \text{Re}[W_{\rm pole}]$ is satisfied
as the coupling strength from the finite volume effect. The acceptable
range is $[1.2 \fm , \, 5.5 \fm]$ for the higher pole of $\Lambda
(1405)$ ($\bar{\mu} _{\bar{K}N} \text{Re}[W_{\rm pole}]=0.53 \gev /
\text{fm}$) and $[2.5 \fm , \, 5.5 \fm]$ for $f_{0}(980)$ ($\bar{\mu}
_{K\bar{K}} \text{Re}[W_{\rm pole}]=0.23 \gev / \text{fm}$).  The
adopted values of $g_{\rm FV}$ are shown as bands in Fig.~\ref{fig:6}
for both $\Lambda (1405)$ and $f_{0}(980)$.

\begin{table}
  \caption{Properties of $\Lambda (1405)$ and $f_{0}(980)$ with finite
    volume effect. Here $\sqrt{\langle r^{2} \rangle}$ is the root mean
    squared distance between two constituent hadrons, while
    $\sqrt{\langle R^{2} \rangle}$ is the root mean squared radius
    evaluated by Eq.~\eqref{eq:Rsize}.}
  \label{tab:4}
  \begin{ruledtabular}
    \begin{tabular*}{8.6cm}{@{\extracolsep{\fill}}cc}
      \multicolumn{2}{c}{$\Lambda (1405)$, higher pole} 
      \\
      \hline
      $B_{\rm E}=m_{N}+m_{K}-\text{Re} [ W_{\rm pole} ]$ & $8.4 \mev$ 
      \\
      $g_{\bar{K} N, \rm FV}$ & $4.6$ -- $5.2 \gev$ 
      \\
      $X_{\bar{K}N,\rm{FV}}$ & $0.82$ -- $1.03$ 
      \\
      $\sqrt{\langle r^{2} \rangle _{\bar{K} N, \rm FV}}$ & $1.7$ -- $1.9 \fm $ 
      \\
      $\sqrt{\langle R^{2} \rangle _{\rm size, FV}}$ & $1.1$ -- $1.2 \fm $ 
      \\
    \end{tabular*}
  \end{ruledtabular}
  \vspace{10pt}
  \begin{ruledtabular}
    \begin{tabular*}{8.6cm}{@{\extracolsep{\fill}}cc}
      \multicolumn{2}{c}{$f_{0}(980)$} 
      \\
      \hline
      $B_{\rm E}=2m_{K}-\text{Re} [ W_{\rm pole} ]$ & $4.2 \mev$ 
      \\
      $g_{K \bar{K}, \rm FV}$ & $2.7$  -- $3.1 \gev$ 
      \\
      $X_{K\bar{K},\rm FV}$ & $0.73$ -- $0.97$
      \\
      $\sqrt{\langle r^{2} \rangle _{K \bar{K}, \rm FV}}$ & $2.6$ -- $3.0 \fm $ 
      \\
      $\sqrt{\langle R^{2} \rangle _{\rm size, FV}}$ & $1.4$ -- $1.6 \fm $
      \\
    \end{tabular*}
  \end{ruledtabular}
\end{table}

The coupling constants and the estimated separations between
constituents are summarized in Table~\ref{tab:4}.  As one can see from
the table, these resonances are dominated by the $\bar{K}N$
($K\bar{K}$) component with large spatial extent. In addition, the
magnitude of the $\bar{K}N$ ($K\bar{K}$) component $X_{\bar{K}N}$
($X_{K\bar{K}}$) is in fair agreement with that obtained on the pole
position presented in Table~\ref{tab:3}.  Root mean squared distances
are $\sqrt{\langle r^{2} \rangle}= 1.7$--$1.9\fm$ for $\Lambda(1405)$
and $2.6$--$3.0 \fm$ for $f_{0}(980)$.  Furthermore, we can estimate
mean squared radii of the resonance states including the finite size
of constituents, $\langle R_{\rm size}^{2} \rangle$, via a relation in
Appendix~\ref{sec:appB}:
\begin{equation}
\langle R^{2}\rangle _{\rm size}
= \frac{m^{2} + m^{\prime 2}}{2 ( m + m^{\prime} )^{2}}
\langle r^{2} \rangle + \frac{1}{2} ( 
\langle x^{2} \rangle 
+ \langle x^{\prime 2} \rangle ) .
\label{eq:Rsize}
\end{equation}
with $\langle x^{( \prime ) 2} \rangle$ being each mean squared radius
of the constituents.  Because $\langle x^{( \prime ) 2} \rangle$ is
just added with a factor $1/2$ to the mean squared radius of the whole
system, size of the constituents always enlarges the mean squared
radius of the system.  Also, due to the kinematic factor $(m^{2} +
m^{\prime 2}) / [2 ( m + m^{\prime} )^{2}]$, if masses of two
constituents for (quasi-)bound system are very similar to each other,
the mean squared distance between constituents $\langle r^{2} \rangle$
corresponds to mean squared diameter rather than radius of the whole
system.  By using the empirical mean squared radii with respect to the
matter distributions for nucleon and kaon estimated from the
electromagnetic radii~\cite{Nakamura:2010zzi},
\begin{equation}
\langle x^{2} \rangle _{N} \approx 0.7 \fm ^{2}, %$ and 
\quad 
\langle x^{2} \rangle _{K, \bar{K}} \approx 0.3 \fm ^{2}, 
\end{equation} 
the root mean squared radii are evaluated as 
\begin{equation}
\sqrt{\langle R^{2} \rangle _{\rm size, FV}} = 
\begin{cases}
1.1 \text{ -- } 1.2 \fm
\quad 
\text{for } \Lambda (1405) , \\
1.4 \text{ -- } 1.6 \fm
\quad 
\text{for } f_{0}(980) ,
\end{cases}
\end{equation}
to which the contributions from the constituent size are about $0.2
\fm$ and $0.1 \fm$ for $\Lambda (1405)$ and $f_{0}(980)$,
respectively.  Both the root mean squared distances and radii for
$\Lambda (1405)$ and $f_{0}(980)$ are larger than the typical hadronic
scale $\lesssim 0.8 \fm$.  In this way, the $\Lambda (1405)$ and
$f_{0}(980)$ can be interpreted as loosely bound Feshbach resonances.

Now we can compare the present result with previous calculations of
the $\bar{K}N$ distance in $\Lambda(1405)$.  In
Ref.~\cite{Sekihara:2010uz} the complex form factors of $\Lambda
(1405)$ was calculated on the pole position in the probe method, and
the (real) mean distance between $\bar{K} N$ was evaluated in the
bound state approximation.  Combining two approaches, we evaluate the
complex ``mean distance'' between $\bar{K} N$ on the pole position at
$1426 - 17 i \mev $ as
\begin{equation}
  \sqrt{\langle r^{2} \rangle_{\bar{K}N}} = 1.22 - 0.63 i \fm \quad 
  |\sqrt{\langle r^{2} \rangle_{\bar{K}N}}| = 1.37 \fm .
\end{equation}
The mean squared distance of $\bar{K} N$ inside $\Lambda (1405)$ on
the pole was also calculated in Ref.~\cite{Dote:2012bu} by using the
complex scaling method with an effective coupled-channel potential.
The result on the pole at $1419 - 14 i \mev $ is
\begin{equation}
  \sqrt{\langle r^{2} \rangle_{\bar{K}N}} = 1.21 - 0.49 i \fm \quad 
  |\sqrt{\langle r^{2} \rangle_{\bar{K}N}}| = 1.31 \fm .
\end{equation}
We find that the estimations of the complex mean distance give a
roughly comparable value with the present result, while the precise
magnitude is about 30--40 \% smaller. In comparison with the real
part, the absolute value is slightly closer to our result, but this is
not a significant difference.  On the other hand, we find that the
present method gives consistent values of the root mean squared
distance evaluated on the real energy axis.  Namely, in
Ref.~\cite{Sekihara:2010uz}, the mean distance of the $\bar{K}N$ bound
system at 1424 MeV was calculated in the probe method, which leads to
\begin{align}
  \sqrt{\langle r^{2}\rangle}
  = & 1.69 \fm 
  \quad (\text{probe method, $B\sim 11$ MeV})
  \nonumber .
\end{align}
In Ref.~\cite{Dote:2008hw}, the mean distance was calculated by the
effective single-channel potential developed in
Ref.~\cite{Hyodo:2007jq} only with its real part, and the results are
\begin{align}
  \sqrt{\langle r^{2}\rangle}
  = & 1.72\text{ -- }1.99 \fm 
  \quad (\text{potential, $B\sim 10$ -- 13 MeV})
  \nonumber .
\end{align}
This consistency is reasonable, because the box size is defined on the
real energy axis in $\tilde{G}(s)$ which is analytically continued to
the complex energy plane to probe behavior of resonance states.

Finally let us discuss uncertainties coming from the choice of the
mass of the resonance states.  Until now we have identified the real
part of the pole position as the ``mass'' of the resonance state,
while the resonance mass may have uncertainties of $\pm \Gamma/2= \mp
\text{Im}[W_{\rm pole}]$.  However, this is a subtle problem, because
the mean distance for a bound state is sensitive to the binding energy
as seen in Eq.~\eqref{eq:rmsfV}.  For a weakly bound state, even a
small variation of the ``mass'' (in particular an upward shift) would
result in a drastic change of the distance.  To assess this
uncertainty, we identify $M_{B}=\text{Re}[W_{\rm pole}] -\Gamma/2$,
with $\Delta M_{B}$ being unchanged, and calculate the closed channel
component $X_{i}$ in addition to the mean distance $\sqrt{\langle
  r^{2}\rangle_{i}}$.  In this case, the compositeness of $\bar{K} N$
and $K \bar{K}$ inside the $\Lambda (1405)$ (higher pole) and
$f_{0}(980)$ are about $0.84$ and $0.75$, respectively.  The root mean
squared distances are calculated as about $1.0 \fm $ for
$\Lambda(1405)$ and $1.1 \fm$ for $f_{0}(980)$.  We see that the
fraction of the closed channel component is obtained within the error
band for the result with $M_{\rm B} = \text{Re}[W_{\rm pole}]$.  On
the other hand, the mean distance for the resonances becomes small,
reflecting the increase of the binding energy.  This means that the
structure of the resonances barely changes with respect to the choice
of the ``mass'', while the mean distance for the states would decrease
when the binding energy is increased.  This analysis indicates the
importance of the precise determination of the pole position of
$\Lambda(1405)$ and $f_{0}(980)$ for the quantitative study of the
spatial structure.

\section{Conclusion}
\label{sec:conclusion}

In this paper the structure of dynamically generated hadrons has been
discussed from the viewpoint of the finite volume effect. We have
presented a method to extract the properties of a bound state in
single-channel scattering using the finite volume mass
shift. Introducing a dynamical scattering model, we have shown that
the coupling strength, compositeness, and mean squared distance
between constituents of the bound state in infinite volume can be
reproduced with good accuracy from the mass shift of the bound state
in finite volume.

This technique has been extended to a quasi-bound state with finite
width in coupled-channel scattering, provided that the width is
small. We can estimate the spatial separation of the components in a
closed channel from the movement of the pole position along with the
finite volume effect on this channel. For an application to physical
resonances, we have considered $\Lambda (1405)$, $\sigma$,
$f_{0}(980)$, and $a_{0}(980)$ described in chiral unitary approach
for coupled-channel hadron scatterings.  Applying the finite volume
effect on the $\bar{K} N$ and $K \bar{K}$ channels, we have found that
the poles for the higher $\Lambda (1405)$ and $f_{0}(980)$ move
downward in finite boxes.  This result indicates that $\Lambda (1405)$
and $f_{0}(980)$ respectively have large $\bar{K}N$ and $K\bar{K}$
components.  Fitting to the mass shift formula, spatial distances of
$\bar{K} N$ and $K \bar{K}$ have been evaluated as $1.7$--$1.9\fm$ and
$2.6$--$3.0 \fm$ for higher $\Lambda (1405)$ and $f_{0}(980)$,
respectively.  Furthermore, with spatial structures of constituents
taken into account, the root mean squared radii of $\Lambda (1405)$
and $f_{0}(980)$ are estimated as $1.1$--$1.2 \fm$ and $1.4$--$1.6
\fm$, respectively, to which the contributions from constituent size
are about $0.2 \fm$ for $\Lambda (1405)$ and $0.1 \fm$ for
$f_{0}(980)$.  Both the root mean squared distances and radii for
$\Lambda (1405)$ and $f_{0}(980)$ are larger than the typical hadronic
scale $\lesssim 0.8 \fm$.

\begin{acknowledgments}
  We acknowledge K.~Sasaki, Y.~Koma, and H.~Suganuma for useful
  discussions.
This work is partly supported by the Grant-in-Aid for Scientific
Research from MEXT and JSPS (No. 22-3389, % Sekihara, Gakushin.
%).
No. 24105702 and No. 24740152) % Hyodo
and by the Global Center of Excellence Program by MEXT, Japan through
the Nanoscience and Quantum Physics Project of the Tokyo Institute of
Technology.

\end{acknowledgments}

\appendix

\section{Mass shift of bound states in finite boxes}
\label{sec:appA}

In this Appendix we derive the leading contribution to the mass shift
formula for bound states in a periodic finite box of the size $L$,
Eq.~\eqref{eq:mass-shift}, following Refs.~\cite{Luscher:1985dn,
  Koma:2004wz}. Here we consider a bound state with mass $M_{\rm B}$
coupled with a two-particle system with masses $m$ and $m^{\prime}\leq
m$. In this Appendix, we work in the Euclidean space. We consider the
small binding region as
\begin{align}
  \sqrt{m^{2}+m^{\prime 2}}
  <& M_{\rm B} < m+m^{\prime} 
  \label{eqA:condition} .
\end{align}

\begin{figure}[!Ht]
  \centering
  \begin{tabular*}{8.6cm}{@{\extracolsep{\fill}}cc}
    \Psfig{4.0cm}{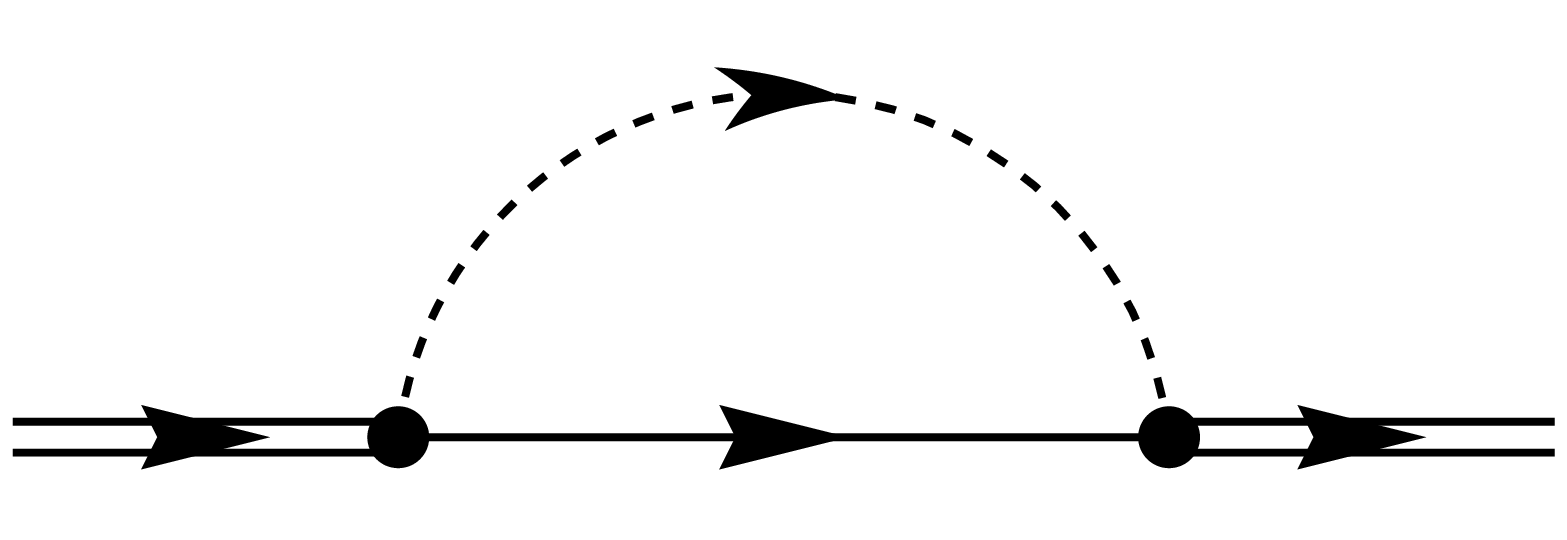} & 
    \Psfig{4.0cm}{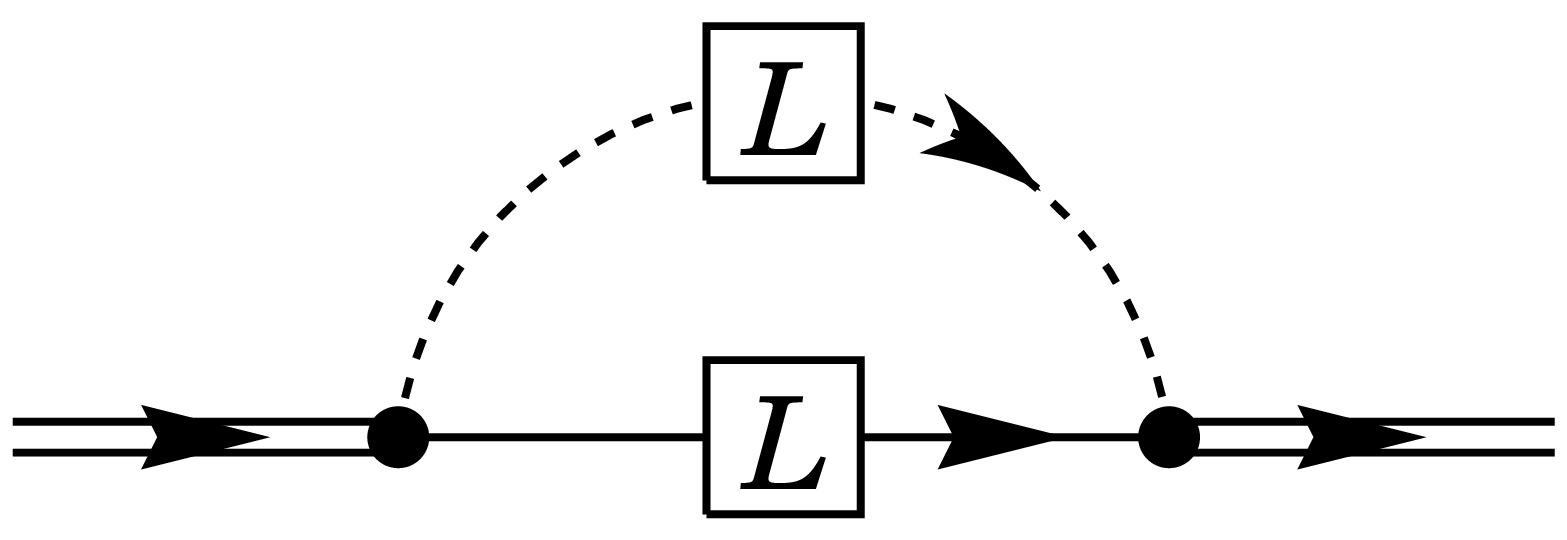} \\
    $\Sigma ( P )$ & $\tilde{\Sigma} ( P )$
  \end{tabular*}
  \caption{Feynman diagrams for the self-energy of the bound state in
    infinite volume (left) and finite volume (right).  The dashed and
    solid lines denote two distinguishable particles and the double
    line stands for the bound state.  Symbol $L$ in the diagram
    represents the finite volume effect.  }
  \label{fig:7}
\end{figure}

In a finite spatial volume, the momentum of the two-particle system is
discretized as $\bm{q}(L) = 2 \pi \bm{n}/L$ with $\bm{n} \in
\mathbb{Z}^{3}$ and the mass of the bound state is shifted to be
$\tilde{M}_{\rm B}(L)\neq M_{\rm B}$. Expanding the self-energy in
finite volume $\tilde{\Sigma} ( P )$ around $P^{0}=iM_{B}$, the mass
shift $\Delta M_{\rm B}(L)\equiv \tilde{M}_{\rm B}(L)-M_{\rm B}$ is
given by
\begin{align}
  \Delta M_{\rm B} ( L ) 
  =& -\frac{1}{2 M_{\rm B}} [ \tilde{\Sigma}  ( P ) - \Sigma ( P ) ] 
  +\mathcal{O}(\Delta M_{\rm B}^{2}) ,
  \label{eqA:DMB}
  \\ 
  P^{\mu} 
  =& ( iM_{\rm B} , \, \bm{0} ) ,
\end{align}
where $\Sigma ( P )$ is the self-energy in the infinite volume. While
several diagrams contribute to the self-energy~\cite{Koma:2004wz}, the
leading effect to the mass shift stems from the diagram shown in
Fig.~\ref{fig:7}. The momentum-discretized loop integral can
be expanded in powers of $e^{-iL\bm{m}\cdot\bm{q}}$, $\bm{m}\in
\mathbb{Z}^{3}$ with the help of the Poisson summation formula, and
the leading contribution can be obtained as
\begin{align}
\tilde{\Sigma} ( P ) - \Sigma ( P ) 
= &\int \frac{d^{4} q}{(2 \pi )^{4}} 
2 \sum _{i=1}^{3} \cos ( L q_{i} ) \nonumber \\
&\times \Gamma
G_{m}[(1-\delta)P+q]G_{m^{\prime}}(\delta P-q)
\Gamma \nonumber \\
&+\mathcal{O}(e^{-\sqrt{2}\bar{\mu} L}) ,
\label{eqA:Gdiff}
\end{align}
where $\Gamma$ is the three-point vertex function, $G_{m}$ is the
propagator with mass $m$, and
\begin{align}
\bar{\mu} 
=& \frac{\sqrt{-\lambda(M_{B}^{2},m^{2},m^{\prime 2})}}
{2M_{B}} ,
\nonumber
\end{align}
with $\lambda (x,\, y,\, z)=x^{2}+y^{2}+z^{2}-2xy-2yz-2zx$.

The momentum fraction $\delta>0$ is chosen to maximize the analytic
region of $\text{Im } q_{1}$ as follows. Here we consider that the
particles of $M_{B}$ and $m$ have a conserved charge\footnote{In the
  application to $\Lambda(1405)$, baryon number is conserved for
  $\Lambda(1405)$ and $N$. For $f_{0}$ in the $K\bar{K}$ scattering,
  we can apply the formula in the equal mass case $m=m^{\prime}$ by
  L\"uscher~\cite{Luscher:1985dn} except for the symmetric factor
  $1/2$, which coincides with Eq.~\eqref{eqA:mass-shift}.} so that we
can trace the line which connect the external $M_{B}$ and $m$
propagators of the vertex function $\Gamma$. We then use the same
argument with Ref.~\cite{Koma:2004wz} to assign the momenta of the
internal lines in the vertex function $\Gamma$. The conditions to
avoid singularity are found to be
\begin{align}
\left(\text{Im}\left\{
(1-\delta)P\pm \frac{1}{2}q
\right\} \right)^{2}&<m^{2}<M_{B}^{2} ,
\nonumber \\
\left(\text{Im}\left\{
\delta P\pm \frac{1}{2}q
\right\} \right)^{2}&<(m^{\prime})^{2} .
\nonumber
\end{align}
By choosing 
\begin{align}
\delta
=\frac{M_{B}^{2}+(m^{\prime})^{2}-m^{2}}{2M_{B}^{2}} ,
\nonumber
\end{align}
the maximum analytic region for $\text{Im }q_{1}$ is obtained as
\begin{align}
0\leq (\text{Im }q_{1})^{2} 
< 4\bar{\mu}^{2},
\nonumber
\end{align}
The poles of the propagators as functions of $q_{1}$ are given by
\begin{align}
\bar{q}_{1}
= &i\sqrt{q_{0}^{2}+q_{\perp}^{2}+\bar{\mu}^{2}+i(2-2\delta)M_{B}q_{0}} 
\label{eqA:q1bar} , \\
\bar{q}_{1}^{\prime}
= &i\sqrt{q_{0}^{2}+q_{\perp}^{2}+\bar{\mu}^{2}-i2\delta M_{B}q_{0}}
\label{eqA:q1pbar} ,
\end{align}
for $G_{m}[(1-\delta)P+q]$ and $G_{m^{\prime}}(\delta P-q)$,
respectively. Modifying the integration contour properly, we obtain
two terms from these poles with the rest contributions being higher
order corrections after the $q_{1}$ integration:
\begin{align}
\tilde{\Sigma} ( P ) - \Sigma ( P ) 
= &I_{1}+I_{1}^{\prime}+\mathcal{O}(e^{-\sqrt{2}\bar{\mu} L}) ,
\end{align}
with
\begin{align}
I_{1}
=& \left.  6i\int \frac{d q_{0}d^{2}q_{\perp}}{(2 \pi )^{3}} 
\frac{\exp (i L q_{1} )}{2q_{1}}  \Gamma
G_{m^{\prime}}(\delta P-q)
\Gamma \right|_{q_{1}=\bar{q}_{1}} , \nonumber \\
I_{1}^{\prime}
=& \left. 6i\int \frac{d q_{0}d^{2}q_{\perp}}{(2 \pi )^{3}} 
\frac{\exp (i L q_{1} )}{2q_{1}}  \Gamma
G_{m}[(1-\delta )P+q] 
\Gamma \right|_{q_{1}=\bar{q}_{1}^{\prime}} \nonumber ,
\end{align}
where we have used the rotational invariance. The remaining
propagators have a pole in the complex $q_{0}$ plane at
\begin{align}
\bar{q}_{0}
= &0 .
\label{eqA:q0}
\end{align}
The leading contribution to the mass shift formula comes from this
pole. To obtain the saddle point expression, we shift the $q_{0}$
integration path from the real axis to $\text{Im }q_{0}=
-i(1-\delta)M_{B}$ ($\text{Im }q_{0}= i\delta M_{B}$) in $I_{1}$
($I_{1}^{\prime}$). The pole contribution from $\bar{q}_{0}$ is picked
up by $I_{1}^{\prime}$ term. The leading contribution is then given by
\begin{align}
I_{1}^{\prime}
=& 6\int \frac{d^{2}q_{\perp}}{(2 \pi )^{3}} 
\frac{\exp (- L \sqrt{\bar{\mu}^{2}+q_{\perp}} )}{2\sqrt{\bar{\mu}^{2}+q_{\perp}}}  
\frac{g^{2} }{2M_{B}}+\mathcal{O}(e^{-\sqrt{2}\bar{\mu} L}) , \nonumber \\
=& \frac{3g^{2}}{4\pi L M_{B}}e^{-\bar{\mu} L}
+\mathcal{O}(e^{-\sqrt{2}\bar{\mu} L}) , 
\end{align}
where the coupling constant $g$ is defined as the vertex function
$\Gamma$ with all the particles being on the mass shell. In this way,
the mass shift formula can be written as
\begin{align}
  \Delta M_{\rm B} ( L ) 
  =& -\frac{3g^{2}}{8\pi L M_{B}^{2}}e^{-\bar{\mu} L}
+\mathcal{O}(e^{-\sqrt{2}\bar{\mu} L}) .
\label{eqA:mass-shift}
\end{align}
This formula recovers Eq.~(3.37) of Ref.~\cite{Luscher:1985dn} with
$m^{\prime}=m$ and with symmetric factor $1/2$ for identical
particles.

Note that to obtain positive $\delta$ we need
\begin{align}
M_{B}^{2}
> &m^{2}-m^{\prime 2} ,
\label{eqA:polecond}
\end{align}
which is guaranteed by Eq.~\eqref{eqA:condition}. All the above
argument can be applied to the mass shift of $m$ ($m^{\prime}$)
through the $m^{\prime}$-$M_{B}$ loop ($m$-$M_{B}$ loop), by replacing
$\{M_{B}\to m, m\to M_{B}, m^{\prime}\to m^{\prime}\}$ ($\{M_{B}\to
m^{\prime}, m\to M_{B}, m^{\prime}\to m\}$).  However, in the small
binding region~\eqref{eqA:condition}, Eq.~\eqref{eqA:polecond} is only
valid for the self-energy of $M_{B}$, so there is no pole contribution
for the self-energies of intermediate particles of $m$ and
$m^{\prime}$. The mass shift of the intermediate particles are then
given by
\begin{align}
  \Delta m
  =& \mathcal{O}(e^{-m^{\prime} L}) \\
  \Delta m^{\prime}
  =& \mathcal{O}(e^{-m L}) 
\end{align}
which do not alter the result~\eqref{eqA:mass-shift}.

\begin{figure}[!t]
  \centering
  \begin{tabular*}{8.6cm}{@{\extracolsep{\fill}}cc}
    % \Psfig{4.0cm}{diag_nucleon_A} & 
    % \Psfig{4.0cm}{diag_nucleon_B} \\
    \includegraphics[scale=0.186]{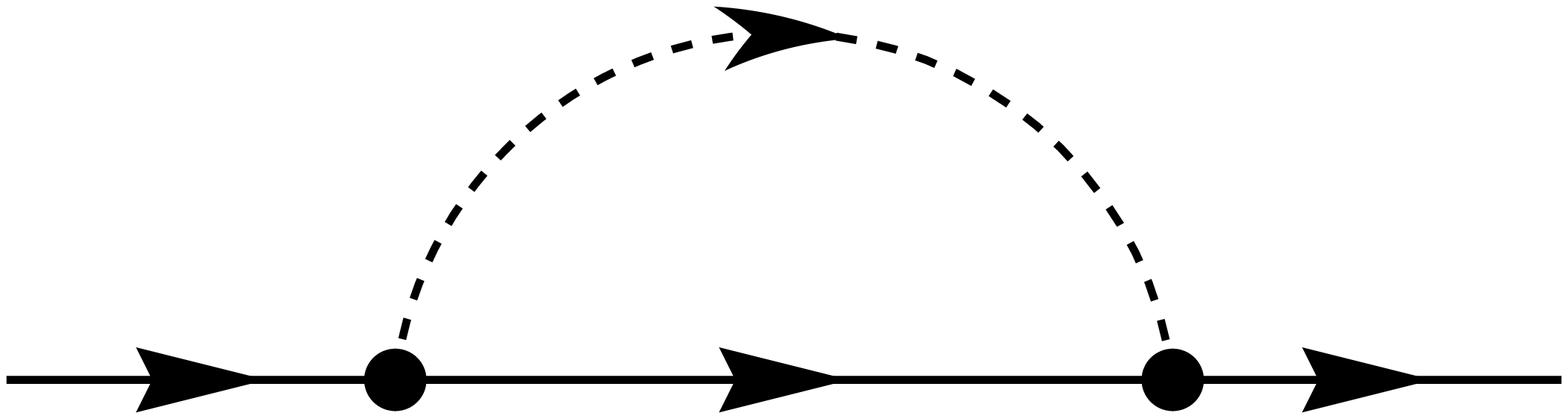} & 
    \includegraphics[scale=0.186]{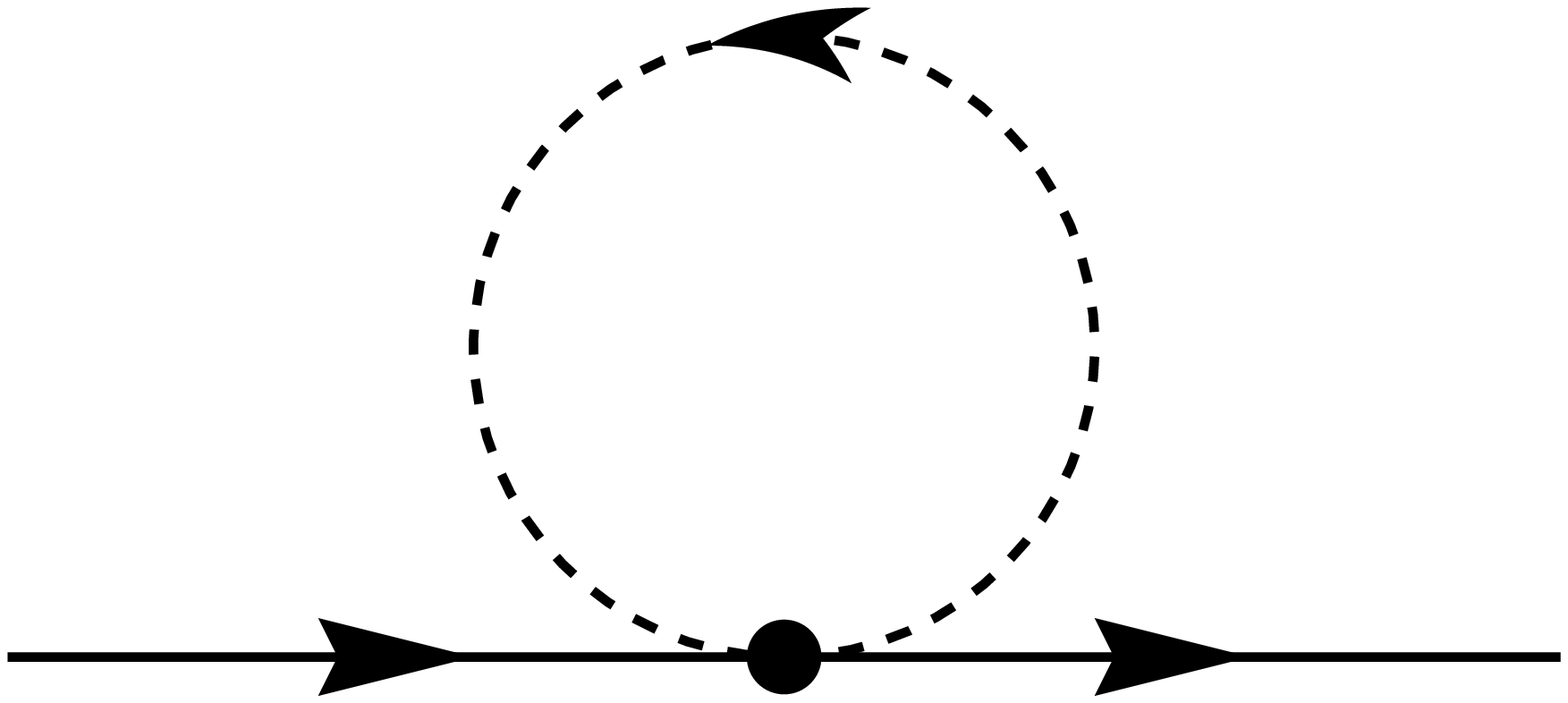} 
  \end{tabular*}
  \caption{Feynman diagrams for the self-energy induced by the
    interaction among constituent particles.  }
  \label{fig:8}
\end{figure}

Finally we consider how mass shift formula~\eqref{eqA:mass-shift} is
modified if the constituent particles have their own spatial
size. This might be crucial to our discussion on dynamically generated
hadronic resonances, because in the real world hadrons have finite
spatial size.  In the present framework, the size of the constituent
particle is induced by the interaction among themselves, which
generates the self-energy diagrams shown in Fig.~\ref{fig:8}. As
studied in Ref.~\cite{Koma:2004wz}, the largest contribution to the
mass shift is
\begin{align*}
  \Delta m 
  =& \mathcal{O} (e ^{- \bar{\mu}^{\prime} L}) , \\
  \quad  \bar{\mu}^{\prime} 
  =& m^{\prime}
  \sqrt{1 - \frac{(m^{\prime})^{2}}{4 m^{2}}}
  =\frac{\sqrt{-\lambda(m^{2}, \, m^{2}, \, m^{\prime 2})}}
  {2m}. 
\end{align*}  
while $\Delta m^{\prime}$ is in higher order than $\Delta m$.  Noting
that $\sqrt{-\lambda(x^{2}, \, m^{2}, \, m^{\prime 2})}/2x$ is a
monotonically decreasing function of $x$ for $\sqrt{m^{2} - m^{\prime
    2}} < x < m + m^{\prime}$, we find $\bar{\mu}<\bar{\mu}^{\prime}$
because of $M_{B}>m$. Again, the mass shift of the constituents is
higher order than the leading contribution of
Eq.~\eqref{eqA:mass-shift}. In general, $\bar{\mu}$ represents the
virtuality of the intermediate state, and the large mass shift is
caused by the channel with small virtuality.
  
In the applications to physical resonances in
Sec.~\ref{sec:resonance}, the finite volume effect is introduced only
to the channel of interest, $\bar{K} N$ or $K \bar{K}$. In some sense,
we use a box which can be felt only by kaons and nucleons, but not by
pions. The spatial structure of hadrons is mainly described by the
pionic cloud, which does not cause the mass shift.  We therefore
conclude that the structure of the constituent hadrons does not alter
the mass shift formula in the practical applications to physical
resonances.  However, the size of the constituent hadrons will modify
the ``size'' of (quasi-)bound states defined by the mean squared
radius, as discussed in Appendix~\ref{sec:appB}.

\section{Relation between size of a bound system and 
distance of constituents inside the system}
\label{sec:appB}

In this Appendix we formulate a relation between mean squared radius
of a dominantly composite two-body bound system and distance of
constituents inside the bound system.  First of all we define
probability that two constituents inside a bound state are in distance
$r$ as $\psi ^{2} (r)$ with the normalization,
\begin{equation}
\int d^{3} r \psi ^{2} ( r ) = 1 .
\end{equation}
Here we assume that the function $\psi ^{2} (r)$ is spherical, {\it
  i.e.}, the two constituents are bound in $s$ wave.  This $\psi ^{2}
(r)$ coincides with the wave function squared with respect to the
relative motion of the two-body bound system, and mean squared
distance, which we have evaluated in a relation to the finite volume
effect, can be evaluated as,
\begin{equation}
\langle r^{2} \rangle = 
\int d^{3} r \, r^{2} \psi ^{2} ( r ) . 
\end{equation}
Next suppose that two constituents, with masses $m$ and $m^{\prime}$,
respectively, have spherical spatial structures of their own.  We
write the density of their spatial structures as $\rho (x)$ and $\rho
^{\prime} (x^{\prime})$, where $x^{(\prime )}$ denotes distance from
the center-of-mass of each constituent, with the normalization,
\begin{equation}
\int d^{3} x \rho ( x ) = 
\int d^{3} x^{\prime} \rho ^{\prime} ( x^{\prime} ) = 1 .
\label{eqB:norm-rho}
\end{equation}
Their own size can be evaluated as the mean squared radii:
\begin{equation}
\langle x^{2} \rangle _{\rm rad}
= 
\int d^{3} x \, x^{2} \rho ( x ) ,
\quad 
\langle x^{\prime 2} \rangle _{\rm rad}
= 
\int d^{3} x^{\prime} \rho ^{\prime} ( x^{\prime} ) .
\end{equation}

Now we can express how one probes matter distribution of the bound
system, in which distance between two constituents is described by
$\psi ^{2} (r)$ and constituents have their own spatial structures
$\rho (x)$ and $\rho ^{\prime} (x^{\prime})$.  Due to the kinematics,
if the relative coordinate of two particles is $\bm{r}$, their
positions measured from the center-of-mass of the bound system can be
expressed as $m^{\prime}\bm{r}/(m+m^{\prime})$ and
$-m\bm{r}/(m+m^{\prime})$, respectively.  Therefore, at position
$\bm{R}$ measured from the center-of-mass of the bound system, the
matter distribution coming from each constituent is expressed as,
\begin{equation}
\sigma ( R ) = \int d^{3} r \psi ^{2} ( r ) 
\rho \left ( \left | \bm{R} - 
\frac{m^{\prime} \bm{r}}{m + m^{\prime}}
\right | \right ) , 
\end{equation}
\begin{equation}
\sigma ^{\prime} ( R ) = \int d^{3} r \psi ^{2} ( r ) 
\rho ^{\prime} \left ( 
\left | \bm{R} + 
\frac{m \bm{r}}{m + m^{\prime}} \right | 
\right ) .
\end{equation}
The normalization of $\sigma (R)$ and $\sigma ^{\prime} (R)$ are found
as,
\begin{equation}
\int d^{3} R \, \sigma ( R ) = \int d^{3} R \, \sigma ^{\prime} ( R ) = 1 ,
\end{equation}
where we have used Eq.~\eqref{eqB:norm-rho} to integrate over $R$.  
In this study we define the whole matter distribution of the bound 
system as an average of the matter distribution coming from the 
two constituents as, 
\begin{equation}
\Rho ( R ) = \frac{1}{2} [ \sigma ( R ) + \sigma ^{\prime} ( R ) ] ,
\end{equation}
with a factor $1/2$ for the correct normalization, 
\begin{equation}
\int d^{3} R \, \Rho ( R ) = 1 .
\end{equation} 
Then the mean squared radius of the bound system, $\langle R^{2}
\rangle _{\rm size}$, can be evaluated as, after simple integral
computations,
\begin{align}
\langle R^{2}\rangle _{\rm size}
& = \int d^{3} R \, R^{2} \Rho ( R ) 
\nonumber \\
& = \frac{m^{2} + m^{\prime 2}}{2 ( m + m^{\prime} )^{2}}
\langle r^{2} \rangle + \frac{1}{2} ( 
\langle x^{2} \rangle _{\rm rad}
+ \langle x^{\prime 2} \rangle _{\rm rad} ) .
\label{eqB:Rsize}
\end{align}
This gives the relation between distance of constituents inside a
bound system and mean squared radius of the whole system.  An
important feature for the mean squared radius of the system is that
each mean squared radius of the constituents is just added with a
factor $1/2$.  If the size of constituents is zero, $\langle x^{2}
\rangle_{\rm rad}=\langle x^{\prime 2} \rangle_{\rm rad}=0$, the mean
squared radius of the bound system corresponds to an average of the
matter distributions coming from two constituents.  The factor $(m^{2}
+ m^{\prime 2}) / [2 ( m + m^{\prime} )^{2}]$ stems from the kinematics.
For example, if the constituent masses are same, $m=m^{\prime}$, the
factor becomes $1/4$, which means that the mean squared distance
between constituents corresponds to, in case that size of constituents
is negligible, mean squared diameter rather than radius of the whole
system.  On the other hand, if one takes $m^{\prime} / m \to 0$ the
factor becomes $1/2$, which can be interpreted as that the mean
squared radius of the whole system is an average of squared distance
$\langle r^{2} \rangle$ coming from the light particle and $0$ from
the heavy particle at the origin.

\end{document}